\documentclass[10pt,twocolumn,twoside]{IEEEtran} 		

\usepackage{cite} 

\ifCLASSINFOpdf \usepackage[pdftex]{graphicx} 
\else 
\fi 

\usepackage{amsmath,amsthm,amssymb}
\newtheorem{theorem}{Theorem}
\newtheorem{proposition}{Proposition}
\newtheorem{corollary}[]{Corollary}

\theoremstyle{remark}
\newtheorem{remark}[theorem]{Remark}

\usepackage{subfig}

\usepackage{enumerate}
\usepackage{bm,makecell}
\usepackage[usenames,dvipsnames]{color}

\newcommand{\vect}[1]{\boldsymbol{#1}}
\newcommand\numberthis{\addtocounter{equation}{1}\tag{\theequation}}

\begin{document} %
	\title{{\huge Accumulate and Jam: Towards Secure Communication via A Wireless-Powered Full-Duplex Jammer}
	\thanks{The authors are with the School of Electrical and Information Engineering, The University of Sydney, NSW 2006, Australia (e-mail: ying.bi@sydney.edu.au, he.chen@sydney.edu.au).}} %
	\author{Ying~Bi and~
	 He~Chen
	 }
	\maketitle
	
	\begin{abstract}
		This paper develops a new cooperative jamming protocol, termed accumulate-and-jam (AnJ), to improve physical layer security in wireless communications. Specifically, a full-duplex (FD) friendly jammer is deployed to secure the direct communication between source and destination in the presence of a passive eavesdropper. We consider the friendly jammer as an energy-constrained node without embedded power supply but with an energy harvesting unit and rechargeable energy storage; it can thus harvest energy from the radio frequency (RF) signals transmitted by the source, accumulate the energy in its battery, and then use this energy to perform cooperative jamming. In the proposed AnJ protocol, based on the energy status of the jammer and the channel state of source-destination link, the system operates in either dedicated energy harvesting (DEH) or opportunistic energy harvesting (OEH) mode. In DEH mode, the source sends dedicated energy-bearing signals and the jammer performs energy harvesting. In OEH mode, the source transmits an information-bearing signal to the destination. Meanwhile, using the harvested energy, the wireless-powered jammer transmits a jamming signal to confound the eavesdropper. Thanks to the FD capability, the jammer also harvests energy from the information-bearing signal that it overhears from the source. We study the complex energy accumulation and consumption procedure at the jammer by considering a practical finite-capacity energy storage, of which the long-term stationary distribution is characterized through applying a discrete-state Markov Chain. An alternative energy storage with infinite capacity is also studied to serve as an upper bound. We further derive closed-form expressions for two secrecy metrics, i.e., secrecy outage probability and probability of positive secrecy capacity. In addition, the impact of imperfect channel state information on the performance of our proposed protocol is also investigated. Numerical results validate all theoretical analyses and reveal the merits of the proposed AnJ protocol over its half-duplex counterpart.
	\end{abstract}
	

	%
	\IEEEpeerreviewmaketitle
			
	\section{Introduction}
		The steady increase and the ubiquity of wireless communications have necessitated an unprecedented awareness of the importance of network security. Unlike the conventional cryptography techniques implemented at higher layers, the physical layer security deals with the properties of physical channels, especially interference and fading, to further strengthen the security of wireless communication systems. Wyner, in his seminal work \cite{wyner_wire-tap_1975}, pioneered the research on physical layer security by pointing out that perfect secrecy can be achieved when the source-eavesdropper channel (i.e. the wiretap channel) is a degraded version of the source-destination channel (i.e. the main channel). Since then, numerous studies attempt to confound eavesdropping from either the perspective of information theory (see \cite{bloch_physical-layer_2011} and references therein) 
 or signal processing (see \cite{mukherjee_principles_2014} for a literature survey). 		

		Artificial noise is one very appealing signal processing approach towards enhanced secrecy \cite{negi_secret_2005}. In \cite{zhou_secure_2010}, a transmitter equipped with multiple antennas is employed to simultaneously transmit information signal to the intended receiver and artificial noise to eavesdroppers. The artificial noise is specifically designed to lie in the null space of the main channel, therefore, only the wiretap channel suffers interference.
		However, this approach becomes inapplicable when the information transmitter has single antenna. To resolve this issue, the concept of cooperative jamming (CJ) was proposed to imitate the effects of multiple transmit antennas; single/multiple helper nodes (now commonly referred to as jammers) work cooperatively with the information transmitter and generate artificial noise to confound the eavesdropper.
		The design and evaluation of CJ schemes for different network setups have attracted a wide range of research interests during the past several years (see \cite{Luo_TIFS_2013_Unc,ng_robust_2014,wang_uncoordinated_2015} for point-to-point wiretap channels, and \cite{dong_improving_2010,Zheng_TSP_2011_Optimal,ding_opportunistic_2011,Chen_CM_2015_Multi,Hoang_TCom_2015_Cooperative} for relay wiretap channels).

		On the other hand, energy deficiency remains to be the bottleneck in the development of ubiquitous wireless communications. Traditionally, this problem is mainly tackled by periodic battery replacement or recharging via gaining energy from various natural energy sources such as solar, wind, and thermal energy. These conventional solutions, however, are restricted by low feasibility and controllability; the battery replacement, in many cases, is often inconvenient (e.g., for a large number of sensors scattered over a wide area) and infeasible (e.g., for sensors implanted inside human bodies). In addition, harvesting energy from natural energy sources is usually climate dependent and thus tends to be intermittent. Recently, wireless energy harvesting (WEH) techniques have drawn much attention as a viable solution to extend the longevity of energy-constrained wireless networks \cite{xiao2014wireless,Chen_TWC_Dist_2015,Chen_TSP_Har_2015,mohammadi_throughput_2016,gu_distributed_2016}. With the WEH technique, wireless communication devices are able to aquire energy from ambient radio frequency (RF) signals, which warrants a new research area of wireless powered CJ.

		In wireless powered CJ schemes, jammers obtain energy from the information source and then use the acquired energy to perform CJ. The effectiveness of a wireless-powered jammer largely depends on its harvested energy. In fact, the intensity of the jamming signal may be brought down to the noise floor if the transmit power is low, whereas a jammer working at a high transmission power needs to reduce its jamming frequency because of the relatively longer charging period. Although finding an optimal jamming power may be one of the compromise solutions to deal with this trade-off, maximizing the energy acquisition is believed to be the approach to solve the problem completely. Of our knowledge, all available studies on CJ with wireless energy harvesting have assumed the jamming nodes to be half-duplex (HD), without investigating the full-duplex (FD) scenarios. In this work, we propose and examine the use of a wireless-powered FD jammer for confounding passive eavesdropping. It mainly brings two merits over its HD counterpart: Firstly, signals sent by the source can always be received by the jammer for scavenging energy, even when the jammer is sending jamming signals. Secondly, the jamming signal can also act as a potential energy source in addition to its original purpose of confusing the eavesdropper. As a consequence, the FD jammer can accumulate more energy and perform more effective jamming than its HD counterpart in the long run. The main contributions of this work are summarized as follows:
		\begin{itemize}
			\item \emph{Protocol design}: This paper explores the use of a wireless-powered FD jammer to secure the communication between a source-destination pair, in the presence of a passive eavesdropper. We consider a time-switched communication protocol with fixed-rate transmission and propose an accumulate-and-jam (AnJ) protocol consisting of dedicated energy harvesting (DEH) and opportunistic energy harvesting (OEH) modes. In DEH mode, the source transfers wireless power to the jammer. In OEH mode, the source sends an information signal to the destination. Being a wireless-powered node, the jammer harvests energy from the RF signals sent by the source, which include the dedicated signal sent to the jammer in DEH mode and the overheard signal sent from the source to the destination in OEH mode. In this way, the benefits of the FD capability are fully exploited to achieve the maximal amount of harvested energy at the jammer. Using the acquired energy, the jammer transmits jamming signals to confound the eavesdropper.
			\item \emph{Imperfect channel state information (CSI)}: The jamming signals sent by the jammer (i.e., artificial noise) are normally designed to lie in the null space of the jammer-destination channel. To accommodate practical limitations of the latest channel estimation techniques, we extend the nulling jamming to allow imperfect CSI at the jammer. Depending on the estimation error factor, this may lead to minor or moderate interference leakage at the destination. Its impact on secrecy performance will be shown via numerical results.
			\item \emph{Finite capacity energy storage}: We have also examined a practical energy storage model with finite capacity at the jammer. Due to the FD mode, the energy storage will experience charging and discharging at the same time. Since a single energy storage is incapable of accommodating this requirement, we adopt a hybrid energy storage system, consisting of one primary energy storage (PES, i.e. a chemical rechargeable battery with high energy density) and one secondary energy storage (SES, i.e. a super-capacitor with high power density), to solve this problem. This hybrid energy storage system is able to be charged and discharged simultaneously. To analyze such a complex procedure, we have applied an energy discretization method and a discrete-state Markov Chain to model the energy state transitions. In addition, energy dissipation caused by, e.g., signal processing and circuitry operation, are also taken into account. Furthermore, an alternative energy storage system with infinite capacity is also studied to serve as a performance upper bound.
			\item \emph{Performance evaluation}: Closed-form expressions for two important metrics, i.e., the secrecy outage probability, and the probability of non-zero secrecy capacity, are derived to evaluate the secrecy performance of our proposed protocol. The secrecy metrics of the wireless-powered HD jammer are also analyzed to provide a benchmark. Numerical results demonstrate that the proposed AnJ protocol substantially outperforms its HD counterpart. Furthermore, design insights regarding the impacts of different system parameters, such as the energy storage capacity, the imperfect CSI, and the antenna allocation at the jammer, are also investigated via numerical results.
		\end{itemize}
		\emph{Relation to Literature}: There are three prior studies relevant to our current work \cite{liu_secure_2015,xing_harvest_2015,krikidis_rf_2012}. In \cite{liu_secure_2015}, the problem of maximizing the secrecy throughput with the aid of a wireless-powered HD jammer was analyzed, with the assumptions that the CSI is perfect, the eavesdropper is noiseless, and the capacity of the jammer's energy storage is infinite. Although our network setup has some similarities with the one used in \cite{liu_secure_2015} in terms of applying a wireless-powered multi-antenna jammer to confound the eavesdropper, the research problem formulation and solutions in our study are essentially different from \cite{liu_secure_2015}. Particularly, the jammer considered in this study operates in FD mode, which leads to mixed operations of energy harvesting and jamming transmission. Moreover, the imperfect CSI, the non-zero receiver noise at the eavesdropper, and the finite energy storage capacity, are also included in our model. In this regard, this work studies the wireless-powered CJ problem with more practical settings than \cite{liu_secure_2015}. In another study of wireless-powered CJ \cite{xing_harvest_2015}, multiple wireless-powered jammers were deployed to secure a two-hop amplify-and-forward relay network. Based on the availability or lack of the eavesdropper's CSI at the transmitter, the upper or lower bounds of the achievable secrecy rate were derived. Nonetheless, the associated energy accumulation process at the jammers was not addressed in \cite{xing_harvest_2015}. In \cite{krikidis_rf_2012}, the transmission outage probability of a wireless-powered relay network was studied. Inspired by the discretized relay battery model proposed in \cite{krikidis_rf_2012}, we also adopt energy discretization and a finite-state MC to model the energy evolvement at the jammer. In addition to the apparently distinct network setup, in order to fulfill the FD operation mode, the jammer studied in this work is equipped with a hybrid energy storage system with the capability of concurrent charging/discharging, whereas the relay in \cite{krikidis_rf_2012} is equipped with a single rechargeable battery which can be in the status of either charging or discharging but not at the same time. As a result, the state transition analysis of the formulated MC in this work is more sophisticated and significantly different from that in \cite{krikidis_rf_2012}. Furthermore, it is also noteworthy that a hybrid energy storage system with infinite capacity is also studied in this work to reveal the performance difference between the discretized and the continuous models.

		\emph{Organization}: Section \ref{sec: system model} describes the system model and the proposed AnJ protocol. Section \ref{sec: markov chain} presents the energy discretization method and the discrete-state Markov Chain for modeling the change of energy status at the jammer during the communication. Closed-form expressions for secrecy outage probability and the existence of non-zero secrecy capacity are derived in Section \ref{sec: performance analysis}. For comparison purposes, two alternative schemes, one with a wireless powered FD jammer with infinite energy storage capacity, the other with a wireless powered HD jammer, are investigated in Section \ref{sub:infinity capacity energy storage} and \ref{sec:HD_jammer}, respectively. After presenting numerical results in Section \ref{sec: simulation}, Section \ref{sec: conclusion} concludes this paper.

		\emph{Notation}: Upper case and lower case bold symbols denote matrices and vectors, respectively. Superscripts $T$ and $\dagger$ represent transposition and conjugate transposition. $\mathrm{Tr}(\vect{A})$ stands for the trace of the matrix $\vect{A}$. $|a|$ is the absolute value of the complex number $a$. $||\vect{a}|| = \sqrt{\vect{a}^\dagger \vect{a}}$ indicates Euclidean norm of the vector $\vect{a}$. $\vect{I}_n$ is the identity matrix of order $n$. $\mathbb{C}^{m\times n}$ denotes the set of all $m \times n$ complex vectors, and $\mathbb{E}\{\cdot\}$ represents the expectation operator.

	\IEEEpubidadjcol
	\begin{figure}[!t]
		\centering
		\includegraphics[width=21pc]{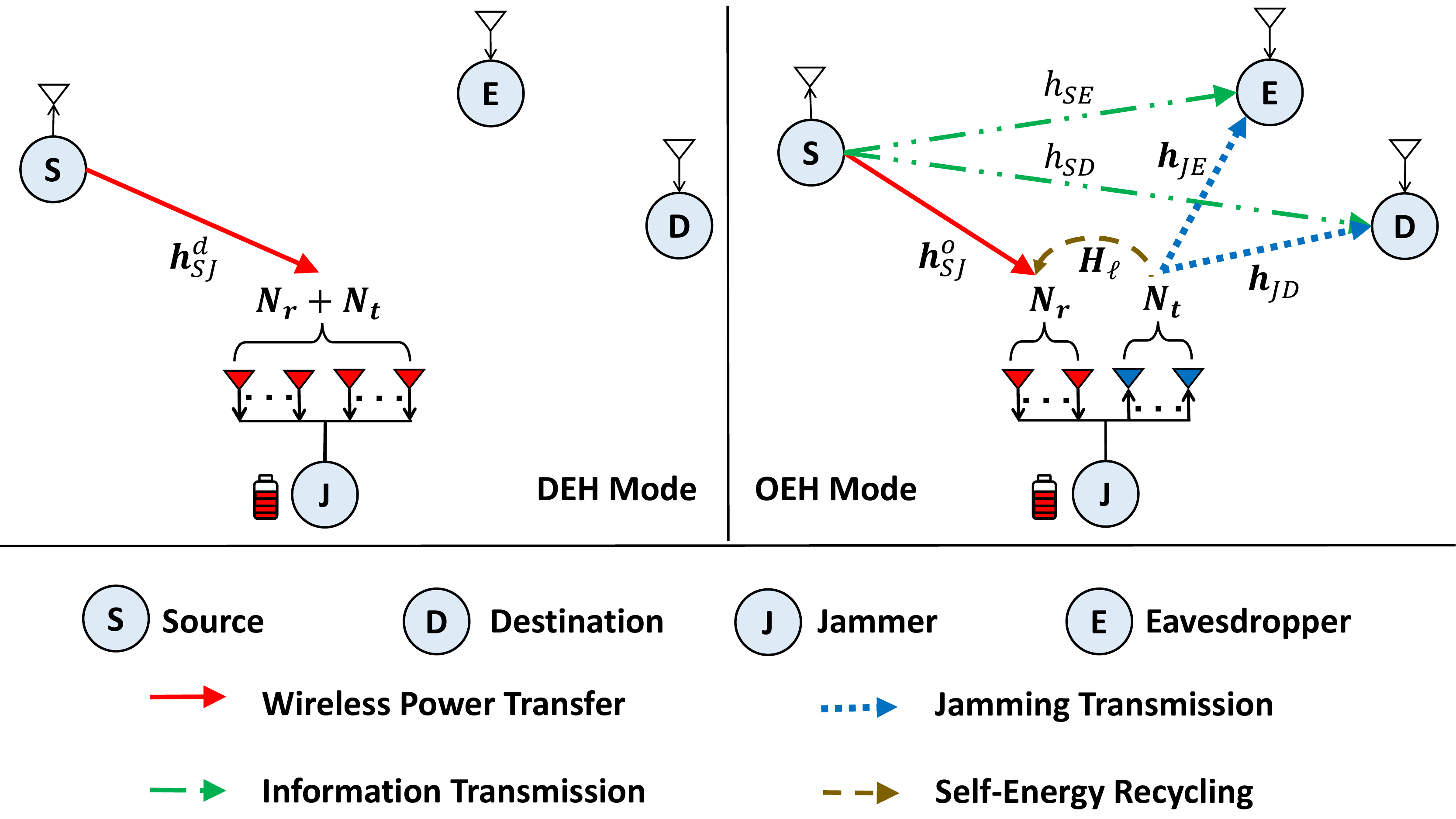} 
		\caption{System model with illustration of the DEH and OEH operation mode.}
		\label{fig_sys_model}
	\end{figure}

	\section{System Model and Proposed Protocol} \label{sec: system model}		
		We consider a point-to-point Gaussian wiretap channel which consists of a source (S), a destination (D), and an eavesdropper (E). Each of these three wireless agents, either being a handset or sensor, is equipped with a single antenna, as illustrated in Fig. \ref{fig_sys_model}.	In line with the vast majority of previous studies, the eavesdropper in this work is considered to be a passive adversary, i.e., it may not transmit but only listens. As such, the instantaneous CSI of E are unknown by any other nodes in the network\footnote{But we assume that the channel distribution of the eavesdropper is available. This assumption has been widely adopted in the literature for secrecy performance analysis \cite{Luo_TIFS_2013_Unc,liu_secure_2015}.}. Moreover, the passive eavesdropping makes the source (S) impossible to determine whether the main channel (S $\to$ D) is superior to the wiretap channel (S $\to$ E).

		\subsection{Jammer Model} \label{sub:definitions_and_assumptions}
			The jammer is assumed to be an energy-constrained node without embedded energy source. It thus needs to acquire energy from ambient RF signals to function. Specifically, when performing cooperative jamming, J operates in FD mode: it uses $N_r$ (i.e., $N_r \ge 1$) antennas to harvest energy from S, and $N_t$ (i.e., $N_t \ge 2$) antennas to transmit jamming signals, simultaneously.
			When cooperative jamming is not carried out, J focuses on energy harvesting with all its antennas, i.e., $N_J = N_t+N_r$ antennas, to receive RF signals. By doing so, J can maximize its energy acquisition by leaving no antennas idle. To enable the aforementioned functionality, J is also equipped with the following components\footnote{Note that at the current stage of research, the optimal structure of an RF energy harvesting node is not completely known. The proposed circuit model in this paper provides only one possible practical design.}:
		\begin{itemize}
			\item $N_J$ RF chains for energy harvesting and jamming transmission,
			\item $N_J$ rectifiers for rectifying the received RF signals into direct currents (DC) \cite{maso_energy-recycling_2015},
			\item a primary energy storage (PES), i.e. a chemical rechargeable battery with high energy density,
			\item a secondary energy storage (SES), i.e. a super-capacitor with high power density.
		\end{itemize}
			Specifically, $N_r$ out of the $N_J$ antennas are connected permanently to $N_r$ rectifiers. The rest ones, i.e., $N_t = N_J - N_r$, are connected to the remaining $N_t$ rectifiers in a non-permanent manner. For simplicity, we consider this antenna allocation as predetermined, and the potential antenna selection problem is beyond the scope of this work.
			The reason for employing the hybrid energy storage system at J is that a single energy storage cannot be charged and discharged at the same time, and therefore cannot support the FD operation. Briefly, PES is directly connected to the rectifiers and the RF chains. When the RF chains are idle, the harvested energy is delivered straight into PES. During transmission, PES uses its stored energy to power up the RF chains. Meanwhile, the harvested energy is temporarily saved in SES. Once the jamming transmission finishes, SES transfers all its stored energy to PES.

			It is also important to clarify that the FD technique applied at J is for simultaneous energy reception and information transmission, which does not strictly fall into the category of conventional FD communications. With a slightly abused terminology, we refer such a jammer as a wireless-powered FD node\footnote{It is also worth noting that such unconventional usage of the term full-duplex is not uncommon in existing literature \cite{zeng_full-duplex_2015,zhong_wireless_2014,mou_exploiting_2015}.}.	In addition, for the comparison purpose, we also consider the usage of a wireless-powered HD jammer, which performs either energy harvesting or jamming transmission, but not at the same time. More details are given in Section \ref{sec:HD_jammer}.

		\subsection{Channel Assumptions} 
			\label{sub:channel_assumptions}
			In Fig. \ref{fig_sys_model}, fading channel coefficients of the links S $\to$ J, S $\to$ D, S $\to$ E, J $\to$ D, and J $\to$ E are denoted by $\vect h_{SJ}$, $h_{SD}$, $h_{SE}$, $\vect h_{JD}$, and $\vect h_{JE}$, respectively. We assume a quasi-static flat fading channel model, in which these fading channel coefficients remain constant within each transmission block of duration $T$\footnote{Without loss of generality, we normalize the block duration to one time unit, i.e., $T=1$. As a consequence, the measures of energy and power become identical in this paper and therefore can be used interchangeably.}, and vary independently from one block to another.
			We apply different small-scale fading models to the channel $\vect h_{SJ}$ that performs energy harvesting and the other channels that perform information transmission. Specifically, since the up-to-date wireless energy harvesting techniques will only work within a relatively short distance, a line-of-sight (LoS) path is likely to present between S and J. Therefore, following \cite{che_multiantenna_2015,xu_multiuser_2015}, we model $\vect h_{SJ}$ as a Rician fading channel. Through adjusting the Rician factor $K$, different channels can be modeled, ranging from a fully deterministic LoS channel (i.e. for short distance) to a weakly dominated LoS channel (i.e. for relatively large distance).
			On the other hand, for all other channels performing information transmission, we apply an independent and identically distributed (i.i.d.) Rayleigh fading to model the heavily scattered wireless communication.

			In this paper, we also make the following assumptions regarding the CSI: $ h_{SD}$ and $\vect h_{JD}$ are acquired respectively by S and J via channel estimation, but the estimated $\vect h_{JD}$ at J is imperfect. The CSI of the eavesdropper, i.e., $h_{SE} $ and $\vect h_{JE} $, is known only to itself as a result of passive eavesdropping. The rationality and advantages of considering imperfect $\vect{h}_{JD} $ at J are two-fold: 1) As an energy-constrained node, the jammer has limited processing power and capability to perform accurate channel estimation, and 2) With imperfect $\vect{h}_{JD} $, we have extended the nulling jamming scheme to allow for interference leakage at the destination. We, therefore, are able to evaluate the impact of imperfect CSI on the system performance. Finally, channel reciprocity is assumed for all the wireless links in the considered system.

	\subsection{Protocol Description} 
		\label{sub:protocol_description}
		In the proposed AnJ protocol, at the beginning of the $k$th transmission block ($k = 1,2,\dots$), J estimates its residual energy $\varepsilon[k]$ and compares it with a predefined threshold $E_{th}$. In the case of $\varepsilon[k] \ge E_{th}$ or $\varepsilon[k] < E_{th}$, J broadcasts a single bit (i.e. $1$ or $0$) to inform S and D whether it is capable of CJ, with bit $0$ indicating that the energy at J is not sufficient for CJ, and bit $1$ otherwise. If bit $0$ is received from J, S feeds back bit $0$ to J and D via a feedback channel to indicate that the current block will operate in DEH mode, and thus D keeps silent during this block. Otherwise (i.e. S and D receives bit $1$ from J), S keeps listening, and D sends a pilot signal for S and J to perform channel estimation. With the assumed channel reciprocity, S can estimate $h_{SD}$ which is then used to verify whether the instantaneous channel capacity $C_{SD}$, expressed as,
		\begin{equation} \label{eq_channel_condition}
			C_{SD} \triangleq \log_2 \left(1+ \frac{P_S H_{SD}}{\sigma_D^2} \right)
		\end{equation}
		can support a secrecy rate $R_s$. In \eqref{eq_channel_condition}, $P_S$ is the source transmitting power, $H_{SD} \triangleq |h_{SD}|^2$, and $\sigma_D^2$ is the additive white Gaussian noise (AWGN) at D. When $C_{SD} \ge R_s$, S feeds back bit $1$ to D and J to indicate that the current block will operate in OEH mode. Otherwise (i.e. $C_{SD} < R_s$), S feeds back bit $0$ to indicate that the DEH mode will be activated. The signaling message exchange among S, D, and J then terminates here. Let $\Phi[k] \in \{\Phi_d,\Phi_o\}$ indicate the operation mode (i.e., either DEH or OEH) for the $k$th transmission block, we have
		\begin{equation} \label{eq_Phi}
			\Phi[k] = \begin{cases}
						\Phi_o, & \text{if } \varepsilon[k] \ge E_{th} \text{ and } C_{SD} \ge R_s, \\
						\Phi_d, & \mbox{otherwise}.
						\end{cases}
		\end{equation}

		The condition of $\varepsilon[k] > E_{th}$ is referred to as the ``\emph{energy condition}''. It is used to prevent the intensity of the jamming signal from dropping down to the noise level at E, which wastes the acquired energy at J. The condition of $C_{SD} > R_s$ is referred to as the ``\emph{channel condition}''. The system will suffer from secrecy outage if the channel condition is not met. The necessity of the channel condition will be shown clearly in Section \ref{sec: performance analysis} where the secrecy outage is defined. 

		In the following, we present the details of the signal processing occurred in each mode.

	\subsubsection{\textbf{In DEH Mode}}
		With a fixed transmitting power $P_S$, S sends J an energy-bearing signal, which is randomly generated and contains no secret information. {To maximize the acquired energy, J employs all the $N_J$ antennas for receiving RF signals.}  By ignoring the negligible energy harvested from the receiver noise, the total amount of energy harvested at J during a DEH block is given by \cite{xing_harvest_2015}
		\begin{equation} \label{eq_Ehd}
			E_h^d = \eta P_S H_{SJ}^d,
		\end{equation}
		where $H_{SJ}^d \triangleq \left\|\vect{h}_{SJ}^d\right\|^{2}, \vect{h}_{SJ}^d \in \mathbb{C}^{N_J\times 1}$, and $\eta$ denotes the energy conversion efficiency. The harvested energy is delivered straight into PES.

	\subsubsection{\textbf{In OEH Mode}}
		With the same transmitting power $P_S$, S sends an information-bearing signal $x_S$ to D, with $\mathbb{E}\{\left|x_S\right|^2\} = 1$. J sends a jamming signal $\vect x_J$, which is deliberately designed for the purpose of producing a null at D and degrading the wiretap channel of E. It is clear that only when J is equipped with $N_t \ge 2$ antennas there are enough degrees of freedom to design $\vect x_J$. Specifically, the artificial noise generation method proposed in \cite{negi_secret_2005} is adopted, which requires instantaneous CSI $\vect h_{JD}$ for beam design\footnote{The jammer may estimate $\vect h_{JD}$ via channel training methods designed specifically for wireless-powered networks, e.g., \cite{xu_energy_2014,zeng_optimized_2015}.}. Unfortunately, as an energy-constrained node, the processing capability of the jammer is limited. Therefore, there is a certain degree of mismatch between the estimated CSI $\bm\hat{\vect h}_{JD}$ and the real CSI $\vect h_{JD} $, of which relation can be expressed as \cite{chen_secrecy_2015,michalopoulos_amplify-and-forward_2012}
		\begin{equation}
			\vect h_{JD} = \sqrt{\rho} \; \bm\hat{\vect h}_{JD} + \sqrt{1-\rho}\; \vect h_{err}
		\end{equation}
		where $\vect h_{err}$ is the error noise vector with i.i.d. zero mean and variance $\sigma_{err}^2$.
		$\rho$, scaling from $0$ to $1$, is the correlation coefficient between $\bm\hat{\vect h}_{JD} $ and $\vect h_{JD}$. A larger $\rho$ means better CSI accuracy. If $\rho = 1$, J has full CSI $\vect h_{JD}$.

		As mentioned earlier, the imperfect estimate $\bm\hat{\vect h}_{JD} $ is used to design the jamming signal $\vect x_J$. Specifically, $\vect x_J = \vect {Wv}$, where $\vect W$ is a $N_t\times (N_t-1)$ matrix constructed in the null-space of $ \hat{\vect h}_{JD} $, and $\vect v$ is the artificial noise vector with  $ N_{t}-1 $ elements. Each element of $\vect v$ is assumed to be an i.i.d. complex Gaussian random variable with zero mean and normalized variance. Thus, the received signal at D and E are given by
		\begin{align*} \label{eq_y_D}
				y_D = \sqrt{P_{S}} h_{SD} x_{S} + \sqrt{(1-\rho)P_J} \vect h_{err}^{\dagger} \dfrac{\vect{Wv}}{\sqrt{N_{t}-1}}+ n_{D} \numberthis
		\end{align*}
		and
		\begin{equation} \label{eq_y_E}
			y_{E} = \sqrt{P_{S}} h_{SE} x_{S} + \sqrt{P_{J}} \vect{h}_{JE}^{\dagger} \dfrac{\vect{Wv}}{\sqrt{N_{t}-1}} + n_{E},
		\end{equation}
		respectively, where $P_J$ is the transmitting power of J (i.e., $0<P_J<E_{th}$). $n_D$ and $n_E$ denote the AWGN with zero mean and variance $\sigma_D^2$ and $\sigma_E^2$, respectively. It can be seen from \eqref{eq_y_D} that the jamming signal also leaks into D's receiver due to the estimation error. We show later at the numerical results in Section \ref{sec: simulation} the impact of $\rho$ on the protocol performance.

		Apart from information transmission and reception, wireless energy harvesting continues in OEH mode. Specifically, the received signal at J is given by
		\begin{equation} \label{eq_y_J}
			y_J = \sqrt{P_{S}} \vect h_{SJ}^o x_{S} + \sqrt{P_J} \vect H_{\ell} \dfrac{\vect{Wv}}{\sqrt{N_{t}-1}}+ \vect n_J
		\end{equation}
		where $\vect H_{\ell} \in \mathbb{C}^{N_r \times N_t}$ denotes the loop channel at J, $\vect n_J$ is the $N_r \times 1$ AWGN vector satisfying $\mathbb{E}[\vect n_J \vect n_J^\dagger] = \vect I_{N_r}\sigma_J^2 $, and $\sigma_J^2$ is the noise variance at each receiving antenna.

		From \eqref{eq_y_J}, the energy-harvesting circuitry of J not only harvests energy from the signal that it overhears from S, but also recycles a portion of its own transmitted energy.
		However, to the best of our knowledge, the distribution of the loop channel $ \vect H_{\ell}$ before applying any interference cancellation is still unknown in open literature. As a result, it is extremely difficult to derive the statistical functions for the recycled energy term. To make the ensuing mathematical analysis tractable and to attain meaningful results, we have to omit the recycled energy term in the following theoretical derivations. In this case, the total amount of harvested energy in OEH mode is expressed as
		\begin{equation} \label{eq_Eho}
			E_h^o \approx \eta P_S H_{SJ}^o
		\end{equation}
		where $H_{SJ}^o \triangleq \left\|\vect{h}_{SJ}^o\right\|^{2}, \vect{h}_{SJ}^o \in \mathbb{C}^{N_r \times 1}$. Similar to \eqref{eq_Ehd}, the harvested noise power is also ignored in \eqref{eq_Eho}. The harvested energy is first saved at SES and then delivered to PES once the jamming transmission finishes. It is noteworthy that due to the omission of recycled energy, strictly speaking, the ensuing theoretical analysis in this paper draws a lower bound for the secrecy performance of the proposed AnJ protocol.

	\section{Markov Chain of Jammer Energy Storage} \label{sec: markov chain}
		The purpose of studying the jammer's energy storage is to find out the probability that the energy condition is met. Due to the FD operation mode, the energy status at J exhibits a complex charging and discharging behavior. We tackle this problem by first applying energy discretization at PES, then using a finite-state Markov Chain (MC) to model the state transition between discrete energy levels.

    \subsection{Energy Discretization} \label{sec_sub: battery discretization}
		As we consider a practical energy storage with finite capacity, the analyses designed for infinite battery capacity \cite{liu_secure_2015} are not applicable to the current study. We, therefore, follow \cite{krikidis_rf_2012} and apply a discrete-level battery model to characterize the dynamic behaviors of PES and SES. Specifically, we discretize PES into $L+1$ energy levels, with the $i$-th level expressed as $ \varepsilon_{i} \triangleq iC_1/L $, $ i \in \{0, 1,\dots , L\} $, where $C_1$ represents the capacity of PES and is assumed to be greater than $E_{th}$ (i.e., otherwise the jammer can never transmit).

		Specifically, in DEH mode, the discretized energy saved in PES can be expressed as
		\begin{equation} \label{eq_ehd}
			\varepsilon_h^d \triangleq \varepsilon_{i_h^d}, \mbox{ where } i_{h}^d = \arg\max_{i \in \{0,1,\dots,L\}}\{\varepsilon_{i}: \varepsilon_{i} \leq  E_h^d \}.
		\end{equation}
		It is worth pointing out that if $E_h^d > C_1$, energy will overflow because the maximum amount of energy that can be saved in PES is $C_1$. Eq. \eqref{eq_ehd} implies $\varepsilon_h^d \le \varepsilon_L = C_1$ by limiting  $i \in \{0,1,\dots,L\}$.
		As for the OEH mode, since the acquired energy $E_h^o$ is first saved in SES and then transferred to PES, considering energy overflow at SES, the amount of energy exported by SES is equal to $\min \{E_h^o, C_2\}$. After transferring along the circuitry from SES to PES, the amount of energy imported to PES can be expressed as
		\begin{equation} \label{eq_Eho_tilde}
			\tilde{E}_h^o = \eta' \times \min \{E_h^o, C_2\}
		\end{equation}
		where $\eta'$ is the energy transfer efficiency from SES to PES \cite{khaligh_battery_2010}, and $\min \{x, y\} $ gives the smaller value between $x$ and $y$. After discretization, the amount of energy eventually saved in PES is given by
		\begin{equation}  \label{eq_eho}
			\varepsilon_h^o \triangleq \varepsilon_{i_h^o}, \mbox{ where } i_{h}^o = \arg\max_{i \in \{0,1,\dots,L\}}\{\varepsilon_{i}: \varepsilon_{i} \le \tilde{E}_h^o \}.
		\end{equation}
		On the other hand, the required energy for jamming transmission $E_{th}$ corresponds to a discrete energy level $\varepsilon_t$, which is defined as
		\begin{align} \label{eq_et}
			\varepsilon_{t} \triangleq \varepsilon_{i_t}, \mbox{ where } i_t = \arg\min_{i \in \{0,1,\dots,L\}}\{\varepsilon_{i}: \varepsilon_{i} \geq E_{th}\}.
		\end{align}
		Note that $E_{th}$ entails all energy consumption occurred at J, i.e., $E_{th} = P_J + P_{c}$, where $P_{c}$ denotes the constant circuitry power \cite{shi_energy_2016}. 
		Furthermore, \eqref{eq_et} can also be expressed as
		\begin{align}
			\varepsilon_t = \left\lceil \frac{E_{th}}{C_1/L} \right\rceil \frac{C_1}{L} = \frac{\tau}{L}C_1 ,
		\end{align}
		where  $\lceil \cdot \rceil $ stands for the ceiling function, and $\tau \triangleq \left\lceil \frac{E_{th}}{C_1/L} \right\rceil $ is defined for notation simplicity.

		At the beginning of the $[k+1]$th transmission block, the residual energy $\varepsilon[k+1]$ is determined by the operation mode $\Phi[k]$ and the residual energy $\varepsilon[k]$ from the $k$th block. Therefore,
		\begin{equation}
			\varepsilon[k+1] =
			\begin{cases}
				\min \{\varepsilon[k] - \varepsilon_t + \varepsilon_h^o, \; C_1 \} & \mbox{if } \Phi[k] = \Phi_{o}, \\
				\min \{ \varepsilon[k] + \varepsilon_h^d, \; C_1 \} & \mbox{if } \Phi[k] = \Phi_{d}.
			\end{cases}
		\end{equation}		

		It is worth noting that the discrete energy model can tightly approximate its continuous counterpart when the number of the discretization level is sufficiently large \cite{Huang_TWC_2008} ($L \to \infty$ corresponds to a continuous energy storage model).
		The impact of $L$ on the secrecy performance of the proposed protocol is presented in Section \ref{sec: simulation}.

	\subsection{Markov Chain} \label{subsec: markov chain}
		With the energy discretization described above, we are able to model the transition of the PES energy states as a finite-state Markov Chain (MC)\footnote{Note that it is not necessary to model the state transition of the SES since it is used only for temporary energy storage in OEH mode.}. We define state $S_i$ as the residual energy of PES $\varepsilon[k]$ being $\varepsilon_i$. The transition probability $p_{i,j}$ represents the probability of a transition from state $S_i$ to state $S_j$. The transitions of the PES energy states have the following six cases:
		
		\subsubsection{PES remains empty ($S_0 \to S_0$)}
		\label{ssub:pes_remains_empty}
			In this case, the energy condition cannot be met. Therefore, the DEH mode is activated. Provided that PES remains empty after recharging, it indicates that the harvested energy during this DEH block is discretized to zero, i.e., $\varepsilon_h^d =\varepsilon_0 = 0$. From the definition given in \eqref{eq_Ehd} and \eqref{eq_ehd}, the condition of $E_h^d = \eta P_S H_{SJ} < \varepsilon_1 = C_1/L $ must remain if the harvested energy is discretized to zero. The transition probability $p_{0,0}$ is thus described as
			\begin{align*} \label{eq_p_00}
				p_{0,0}  	= & \Pr \left\{ \varepsilon_h^d = 0 \right\} =  \Pr \left\{ E_h^d < \varepsilon_1 \right\} \\
							= & \Pr \left\{ H_{SJ}^d < \frac{1}{\eta P_S L/C_1} \right\} \numberthis
			\end{align*}
			Since the channel between S and J is assumed to be Rician fading, the CDF of $H_{SJ}^d$ is given by \cite{ko_average_2000}
			\begin{equation} \label{eq_CDF_HSJd}
				F_{H_{SJ}^d}(x) = 1 - Q_{N_J} \left( \sqrt{2N_J K}, \sqrt{ \frac{2(K+1)}{\Omega_{SJ}}x }\right),
			\end{equation} where $Q_{N_J}(\cdot, \cdot)$ is the generalized ($N_J$-th order) Macum $ Q $-function \cite{gradshtein_table_2007}, and $K$ is the rician factor. Combining \eqref{eq_CDF_HSJd} with \eqref{eq_p_00}, we have
			\begin{equation} \label{eq_p_00_final}
				p_{0,0} = F_{H_{SJ}^d} \left(\frac{1}{\eta P_S L/C_1}\right)
			\end{equation}

		\subsubsection{PES remains full ($S_L \to S_L$)} 
		\label{ssub:pes_remains_full_}
			In this case, the energy condition is certainly met. The selection of the operation mode thus depends purely on the channel condition. If $C_{SD} \ge R_s$, OEH is invoked, where the consumed energy $\varepsilon_t$ is no larger than the harvested energy $\varepsilon_h^o$. Otherwise (i.e. $C_{SD} < R_s $), DEH is invoked, and the harvested energy $\varepsilon_h^d$ can be any arbitrary value as PES is full and cannot accept more energy. The corresponding transition probability is calculated as
			\begin{align*}
				p_{L,L}  = & \Pr \left\{ C_{SD} \ge R_s \right\} \Pr \left\{ \varepsilon_h^o \ge \varepsilon_t\right\} \\
						   & + \Pr \left\{C_{SD} < R_s \right\} \numberthis
			\end{align*}
			We first evaluate $q_c$, i.e. the probability of the channel condition being satisfied. After performing some simple manipulations, we have
			\begin{align}
				q_c = \Pr \left\{ C_{SD} \ge R_s \right\} = \Pr \left\{H_{SD} \ge \frac{2^{R_s}-1}{P_S/\sigma_D^2} \right \}
			\end{align}
			Since the channel $h_{SD}$ is assumed to be Rayleigh fading, $H_{SD}$ follows an exponential distribution with CDF
			\begin{equation} \label{eq_CDF_HSD}
				F_{H_{SD}}(x) = 1 - \exp \left(- \frac{x}{\Omega_{SD}} \right)
			\end{equation}
			Consequently, the probability for the channel condition is given by
			\begin{equation} \label{eq_channel_ready}
				q_c = 1 - F_{H_{SD}} \left( \frac{2^{R_s}-1}{P_S/\sigma_D^2} \right)
			\end{equation}
			and
			\begin{equation} \label{eq_channel_not_ready}
				\Pr \left\{ C_{SD} < R_s \right\} = 1 - q_c = F_{H_{SD}} \left( \frac{2^{R_s}-1}{P_S/\sigma_D^2} \right)
			\end{equation}

			We now analyze the term $\Pr \left\{ \varepsilon_h^o \ge \varepsilon_t\right\}$. From the definition given in \eqref{eq_Eho_tilde}, we have
			\begin{align*} \label{eq_eho_ge_et}
				\Pr  \left\{ \varepsilon_h^o \ge \varepsilon_t\right\}
				= & \Pr \left\{ (\eta' E_h^o \geq \varepsilon_t) \cap ( E_h^o < C_2)\right\} \\
						& + \Pr \left\{(\eta' C_2 \geq \varepsilon_t) \cap ( E_h^o \geq C_2) \right\} \\
						= & \begin{cases}
								\Pr \{\eta' E_h^o \ge \varepsilon_t\} & \mbox{if } C_2 \ge \frac{\tau}{\eta' L}C_1 \\
								0 								& \mbox{otherwise} \numberthis
							\end{cases}
			\end{align*}
			With the definition of $E_h^o$ given in \eqref{eq_Eho}, we can obtain
			\begin{align*}
				&\Pr \{\eta' E_h^o \ge \varepsilon_t\} = \Pr \left\{ H_{SJ}^o \ge \frac{\tau}{\eta \eta' P_S L / C_1} \right\} \numberthis
			\end{align*}
			Similar to \eqref{eq_CDF_HSJd}, the CDF of $H_{SJ}^o$ is
			\begin{equation} \label{eq_F_HSJ_o}
				F_{H_{SJ}^o}(x) = 1 - Q_{N_r} \left( \sqrt{2N_r K}, \sqrt{ \frac{2(K+1)}{\Omega_{SJ}}x }\right)
			\end{equation}
			Consequently,
			\begin{align*} \label{eq_EHO_ge_et}
				\Pr \{\eta' E_h^o \ge \varepsilon_t\}  = 1 - F_{H_{SJ}^o} \left( \frac{\tau}{\eta \eta' P_S L / C_1} \right) \numberthis
			\end{align*}

			By combining \eqref{eq_channel_ready}, \eqref{eq_channel_not_ready}, \eqref{eq_eho_ge_et}, and \eqref{eq_EHO_ge_et}, we can obtain the transition probability $p_{L,L}$ as
			\begin{align*}
				p_{L,L}
				 = \begin{cases}
						q_c \left(1 - F_{H_{SJ}^o}\left(\frac{\tau}{\eta \eta' P_S L / C_1} \right) \right)  & \mbox{if } C_2 \ge \frac{\tau}{\eta' L}C_1 \\
						F_{H_{SD}}\left(\frac{2^{R_s}-1}{P_S/\sigma_D^2}\right) & \mbox{otherwise} \numberthis
					\end{cases}
			\end{align*}

		\subsubsection{The non-empty and non-full PES remains unchangeable ($S_i \to S_i: 0 < i < L$)} 
			\label{ssub:the_non_empty_and_non_full_pes_remains_unchangeable_}
			In this transition case, we need to first evaluate the energy condition. If the available energy is less than the required energy threshold, i.e., $\varepsilon_i < \varepsilon_t$, DEH mode is invoked. If $\varepsilon_i \ge \varepsilon_t$, then the energy condition is met. Next, we need to evaluate the channel condition. In the case that the channel condition is not satisfied, i.e.,  $C_{SD} < R_S$, again DEH mode is selected. Similar to the first transition probability (i.e., $S_0 \to S_0$) , the unchangeable state transition in DEH mode indicates that the harvested energy is discretized to zero, i.e., $\varepsilon_h^d = 0$. On the other hand, if $\varepsilon_i \ge \varepsilon_t$ and $C_{SD} \ge R_S$ are both satisfied, OEH is activated. The unchangeable state transition in OEH mode indicates that the amount of harvested energy must equal the amount of the consumed energy, i.e., $\varepsilon_h^o = \varepsilon_t$. The transition probability is thus described as
			\begin{align*}
				p_{i,i} = & \Pr \{\varepsilon_i < \varepsilon_t\} \Pr \{\varepsilon_h^d = 0\} \\
						& + \Pr \{\varepsilon_i \ge \varepsilon_t\} \Pr\{C_{SD} < R_s \} \Pr \{\varepsilon_h^d = 0\} \\
						& + \Pr \{\varepsilon_i \ge \varepsilon_t\} \Pr\{C_{SD} \ge R_s \} \Pr \{\varepsilon_h^o = \varepsilon_t\} \\
						= & \begin{cases}
							\Pr \{\varepsilon_h^d = 0\} & \mbox{if } i < \tau \numberthis \\
							\Pr\{C_{SD} < R_s \} \Pr \{\varepsilon_h^d = 0\} & \mbox{if } i \ge \tau \\
							\quad + \Pr\{C_{SD} \ge R_s \} \Pr \{\varepsilon_h^o = \varepsilon_t\}
						\end{cases}
			\end{align*}
			From the definition given in \eqref{eq_Eho_tilde}, we have
			\begin{align*} \label{eq_eho_equal_et}
				\Pr & \{\varepsilon_h^o =  \varepsilon_t\} \\
				= & \Pr \{(0 \le \eta' E_h^o - \varepsilon_t < \varepsilon_1) \cap (E_h^o < C_2) \} \\
				& + \Pr \{ (0 \le \eta' C_2 - \varepsilon_t < \varepsilon_1) \cap (E_h^o \ge C_2) \} \\
				= & \begin{cases}
						0 									& \mbox{if } C_2 < \frac{\tau}{\eta' L}C_1 \\
						\Pr \{ \eta' E_h^o \ge \varepsilon_t \} 	& \mbox{if } \frac{\tau}{\eta' L}C_1 \le C_2 < \frac{\tau+1}{\eta' L}C_1 \\
						\Pr \{ \varepsilon_t \le \eta' E_h^o < \varepsilon_1+\varepsilon_t \} & \mbox{if } C_2 > \frac{\tau+1}{\eta' L}C_1 \numberthis
					\end{cases}
			\end{align*}

			With some simple manipulations, after combining \eqref{eq_p_00}, \eqref{eq_p_00_final}, \eqref{eq_channel_ready}, \eqref{eq_channel_not_ready}, \eqref{eq_EHO_ge_et} and \eqref{eq_eho_equal_et}, the transition probability $p_{i,i}$ is given in \eqref{eq_p_ii} on the top of the next page.


			\newcounter{MYtempeqncnt_p_ii}
			\begin{figure*}[!t]
			\normalsize
			\setcounter{MYtempeqncnt_p_ii}{\value{equation}}
			\setcounter{equation}{29}
			\begin{align*} \label{eq_p_ii}
				p_{i,i}
				= \begin{cases}
						F_{H_{SJ}^d} \left(\frac{1}{\eta P_S L / C_1}\right) & \mbox{if } i < \tau \\
						(1 - q_c) F_{H_{SJ}^d}\left(\frac{1}{\eta P_S L / C_1}\right) 																								 & \mbox{if } i \ge \tau \;\&\; C_2 < \frac{\tau}{\eta' L}C_1 \\
						(1 - q_c) F_{H_{SJ}^d}\left(\frac{1}{\eta P_S L / C_1}\right) + q_c \left(1 - F_{H_{SJ}^o} \left( \frac{\tau}{\eta \eta' P_S L / C_1} \right) \right)& \mbox{if } i \ge \tau \;\&\; \frac{\tau}{\eta' L}C_1 \le C_2 < \frac{\tau+1}{\eta' L}C_1 \\
						(1 - q_c)  F_{H_{SJ}^d}\left(\frac{1}{\eta P_S L / C_1}\right) + q_c \left(F_{H_{SJ}^o} \left( \frac{\tau+1}{\eta \eta' P_S L / C_1}  \right) - F_{H_{SJ}^o} \left( \frac{\tau}{\eta \eta' P_S L / C_1} \right) \right) & \mbox{if } i \ge \tau \;\&\; C_2 \ge \frac{\tau+1}{\eta' L}C_1  \numberthis
					\end{cases}
			\end{align*}
			\setcounter{equation}{\value{MYtempeqncnt_p_ii}}
			\hrulefill
			\vspace*{4pt}
			\end{figure*}

			\newcounter{MYtempeqncnt_p_iL}
			\begin{figure*}[!t]
			\normalsize
			\setcounter{MYtempeqncnt_p_iL}{\value{equation}}
			\setcounter{equation}{33}
			\begin{align*} \label{eq_p_iL}
				p_{i,L}
				= \begin{cases}
					1 - F_{H_{SJ}^d} \left( \frac{L-i}{\eta P_S L / C_1} \right)			& \qquad \qquad \quad \mbox{if } i < \tau \\
					(1 - q_c)  \left(1 - F_{H_{SJ}^d} \left( \frac{L-i}{\eta P_S L / C_1} \right) \right)	& \qquad \qquad \quad \mbox{if } i \ge \tau \;\&\; C_2 < \frac{L-i+\tau}{\eta'L}C_1 \\
					(1 - q_c)  \left(1 - F_{H_{SJ}^d} \left( \frac{L-i}{\eta P_S L / C_1} \right) \right) + q_c \left(1 - F_{H_{SJ}^o}\left( \frac{L-i+\tau}{\eta \eta' P_S L / C_1}  \right) \right)							&  \qquad \qquad \quad  \mbox{if } i \ge \tau \;\&\; C_2 \ge \frac{L-i+\tau}{\eta'L}C_1   \numberthis
				  \end{cases}
			\end{align*}
			\setcounter{equation}{\value{MYtempeqncnt_p_iL}}
			\hrulefill
			\vspace*{4pt}
			\end{figure*}

			\newcounter{MYtempeqncnt_p_ij}
			\begin{figure*}[!t]
			\normalsize
			\setcounter{MYtempeqncnt_p_ij}{\value{equation}}
			\setcounter{equation}{37}
			\begin{align*} \label{eq_p_ij}
			p_{i,j}
			= \begin{cases}
				F_{H_{SJ}^d} \left(\frac{j-i+1}{\eta P_S L / C_1}\right) - F_{H_{SJ}^d} \left(\frac{j-i}{\eta P_S L / C_1}\right) & \qquad \qquad \quad \mbox{if } i < \tau \\
				(1 - q_c)  \left(F_{H_{SJ}^d} \left(\frac{j-i+1}{\eta P_S L / C_1}\right) - F_{H_{SJ}^d} \left(\frac{j-i}{\eta P_S L / C_1}\right) \right) 							& \qquad \qquad \quad \mbox{if } i \ge \tau \;\&\; C_2 < \frac{j-i+\tau}{\eta'L}C_1 \\
				(1 - q_c)  \left(F_{H_{SJ}^d} \left(\frac{j-i+1}{\eta P_S L / C_1}\right) - F_{H_{SJ}^d} \left(\frac{j-i}{\eta P_S L / C_1}\right) \right) 	& \qquad \quad \qquad \mbox{if } i \ge \tau \;\&\; \frac{j-i+\tau}{\eta' L}C_1 \le C_2 < \frac{j-i+\tau+1}{\eta' L}C_1 \\
				\quad + q_c \left(1 - F_{H_{SJ}^o} \left( \frac{j-i+\tau}{\eta \eta' P_S L / C_1}  \right) \right) 	 \\
				(1 - q_c)  \left(F_{H_{SJ}^d} \left(\frac{j-i+1}{\eta P_S L / C_1}\right) - F_{H_{SJ}^d} \left(\frac{j-i}{\eta P_S L / C_1}\right) \right) 	& \qquad \qquad \quad \mbox{if } i \ge \tau \;\&\; C_2 > \frac{j-i+\tau+1}{\eta'L}C_1 \\
				\quad + q_c \left(F_{H_{SJ}^o} \left( \frac{j-i+\tau+1}{\eta \eta' P_S L / C_1} \right) - F_{H_{SJ}^o} \left( \frac{j-i+\tau}{\eta \eta' P_S L / C_1} \right) \right)  \numberthis
			  \end{cases}
			\end{align*}
			\setcounter{equation}{\value{MYtempeqncnt_p_ij}}
			\hrulefill
			\vspace*{4pt}
			\end{figure*}

			\newcounter{MYtempeqncnt_p_ji}
			\begin{figure*}[!t]
			\normalsize
			\setcounter{MYtempeqncnt_p_ji}{\value{equation}}
			\setcounter{equation}{40}
			\begin{align*} \label{eq_p_ji}
				p_{j,i}
				= \begin{cases}
					0 			& \qquad \qquad \qquad \quad \mbox{if } j < \tau \mbox{ or }C_2 < \frac{\tau-j+i}{\eta'L}C_1 \\
					q_c \left(1 - F_{H_{SJ}^o}\left( \frac{\tau-j+i}{\eta \eta' P_S L / C_1} \right) \right)		& \qquad \qquad \qquad \quad \mbox{if } j \ge \tau \;\&\; \frac{\tau-j+i}{\eta'L}C_1 \le C_2 < \frac{\tau-j+i+1}{\eta'L}C_1 \\
					q_c \left(F_{H_{SJ}^o}\left( \frac{\tau-j+i+1}{\eta \eta' P_S L / C_1}  \right) - F_{H_{SJ}^o}\left( \frac{\tau-j+i}{\eta \eta' P_S L / C_1}  \right) \right) 							 & \qquad \qquad \qquad \quad \mbox{if } j \ge \tau \;\&\; C_2 \ge \frac{\tau-j+i+1}{\eta'L}C_1 \numberthis
					\end{cases}
			\end{align*}
			\setcounter{equation}{\value{MYtempeqncnt_p_ji}}
			\hrulefill
			\vspace*{4pt}
			\end{figure*}

		\subsubsection{PES is fully charged  ($S_i \to S_L: 0 \le i < L$)} 
			\label{ssub:pes_is_partially_charged_}
			In this case, the harvested energy after discretization satisfies $\varepsilon_h^d \ge \varepsilon_L - \varepsilon_i$ in DEH, or $\varepsilon_h^o - \varepsilon_t \ge \varepsilon_L - \varepsilon_i$ in OEH. The transition probability $p_{i,L}$ is thus described as
			\setcounter{equation}{30}
			\begin{align*}
				& p_{i,L} \\
				& = \Pr \{\varepsilon_i < \varepsilon_t\} \Pr \{\varepsilon_h^d \ge \varepsilon_L - \varepsilon_i\} \\
						& \quad + \Pr \{\varepsilon_i \ge \varepsilon_t\} \Pr\{C_{SD} < R_s \} \Pr \{\varepsilon_h^d \ge \varepsilon_L - \varepsilon_i\} \\
						& \quad + \Pr \{\varepsilon_i \ge \varepsilon_t\} \Pr\{C_{SD} \ge R_s \} \Pr \{\varepsilon_h^o - \varepsilon_t \ge \varepsilon_L - \varepsilon_i\} \\
						& = \begin{cases}
							\Pr \{\varepsilon_h^d \ge \varepsilon_L - \varepsilon_i\} & \mbox{if } i < \tau \numberthis \\
							(1 - q_c) \Pr \{\varepsilon_h^d \ge \varepsilon_L - \varepsilon_i\} & \mbox{if } i \ge \tau \\
							\; + q_c \Pr \{\varepsilon_h^o - \varepsilon_t \ge \varepsilon_L - \varepsilon_i\}
						\end{cases}
			\end{align*}	
			Particularly, we have
			\begin{align*}
				\Pr \{\varepsilon_h^d \ge \varepsilon_L - \varepsilon_i\}
				& = \Pr \{E_h^d \ge C_1 - \varepsilon_i\} \numberthis \\
			\end{align*}
			and
			\begin{align*}
				\Pr & \{\varepsilon_h^o - \varepsilon_t \ge \varepsilon_L - \varepsilon_i\} \\
				& = \Pr \{ (\eta' E_h^o - \varepsilon_t \ge C_1 - \varepsilon_i ) \cap (E_h^o < C_2)  \} \\
				& \quad + \Pr \{ (\eta' C_2 - \varepsilon_t \ge C_1 - \varepsilon_i ) \cap (E_h^o \ge C_2)  \} \\
				& = \begin{cases}
					0 		& \mbox{if } C_2 < \frac{L-i+\tau}{\eta' L}C_1 \\
					\Pr \{\eta' E_h^o \ge C_1 -\varepsilon_i + \varepsilon_t\} & \mbox{if } C_2 \ge \frac{L-i+\tau}{\eta' L}C_1 \numberthis
					\end{cases} \\
			\end{align*}
			Consequently, the transition probability in this case is given by \eqref{eq_p_iL}.

			\subsubsection{PES is partially charged  ($S_i \to S_j: 0 \le i < j < L$)} 
			\label{ssub:pes_is_partially_charged_}
			This transition case can happen either in DEH mode with $\varepsilon_h^d = \varepsilon_j - \varepsilon_i$, or in OEH mode with $\varepsilon_h^o - \varepsilon_t = \varepsilon_j - \varepsilon_i$. The transition probability $p_{i,j}$ is therefore calculated as
			\setcounter{equation}{34}
			\begin{align*}
				p_{i,j}
				= & \Pr \{\varepsilon_i < \varepsilon_t\} \Pr \{\varepsilon_h^d = \varepsilon_j - \varepsilon_i\} \\
					& + \Pr \{\varepsilon_i \ge \varepsilon_t\} \Pr\{C_{SD} < R_s \} \Pr \{\varepsilon_h^d = \varepsilon_j - \varepsilon_i\} \\
					& + \Pr \{\varepsilon_i \ge \varepsilon_t\} \Pr\{C_{SD} \ge R_s \} \Pr \{\varepsilon_h^o - \varepsilon_t = \varepsilon_j - \varepsilon_i\} \\
					= & \begin{cases}
						\Pr \{\varepsilon_h^d = \varepsilon_j - \varepsilon_i\} & \mbox{if } i < \tau \numberthis \\
						(1 - q_c) \Pr \{\varepsilon_h^d = \varepsilon_j - \varepsilon_i\} & \mbox{if } i \ge \tau \\
						\quad + q_c \Pr \{\varepsilon_h^o - \varepsilon_t = \varepsilon_j - \varepsilon_i\}
						\end{cases}
			\end{align*}
			Specifically, we have
			\begin{align*}
				\Pr  \{\varepsilon_h^d = \varepsilon_j - \varepsilon_i\} = \Pr \{\varepsilon_j - \varepsilon_i \le \varepsilon_h^d < \varepsilon_{j+1} - \varepsilon_i\} \numberthis \\
			\end{align*}
			and
			\begin{align*}
				\Pr & \{\varepsilon_h^o - \varepsilon_t = \varepsilon_j - \varepsilon_i\} \\
				= & \Pr \{(\varepsilon_j - \varepsilon_i \le \eta' E_h^o - \varepsilon_t < \varepsilon_{j+1} - \varepsilon_i) \cap (E_h^o < C_2) \} \\
				& + \Pr \{ (\varepsilon_j - \varepsilon_i \le \eta' C_2 - \varepsilon_t < \varepsilon_{j+1} - \varepsilon_i) \cap (E_h^o \ge C_2) \} \\
				= & \begin{cases}
						0 									\qquad \qquad \quad \mbox{if } C_2 < \frac{j-i+\tau}{\eta' L}C_1 \\
						\Pr \{ \eta' E_h^o \ge \varepsilon_j - \varepsilon_i +  \varepsilon_t \} 	 \\
						\qquad \qquad \quad  \mbox{if } \frac{j-i+\tau}{\eta' L}C_1 \le C_2 < \frac{j-i+\tau+1}{\eta' L}C_1  \numberthis \\
						\Pr \{ \varepsilon_j - \varepsilon_i + \varepsilon_t \le \eta' E_h^o < \varepsilon_{j+1} - \varepsilon_i +\varepsilon_t \} \\
						 \qquad \qquad \quad  \mbox{if } C_2 > \frac{j-i+\tau+1}{\eta' L}C_1
					\end{cases} \\
			\end{align*}
			Therefore, we can obtain the transition probability in this case given by \eqref{eq_p_ij} shown on top of the page.

		\subsubsection{PES is discharged  ($S_j \to S_i: 0 \le i < j \le L$)} 
			\label{ssub:pes_is_discharged_}
			Since the stored energy is reduced during this transition case, the OEH operation mode must have been activated. The amount of reduced energy, i.e., $\varepsilon_j - \varepsilon_i$, equals the difference between the consumed energy $\varepsilon_t$ and the discretized acquired energy $\varepsilon_h^o$. The transition probability is expressed as
			\setcounter{equation}{38}
			\begin{align*}
				p_{j,i}
				& = \Pr \{\varepsilon_j \ge \varepsilon_t\} \Pr\{C_{SD} \ge R_s \} \Pr \{\varepsilon_t - \varepsilon_h^o  = \varepsilon_j - \varepsilon_i\} \\
				& = \begin{cases}
					0 		& \mbox{if } j < \tau \\
					q_c \Pr \{\varepsilon_t - \varepsilon_h^o  = \varepsilon_j - \varepsilon_i\} & \mbox{if } j \ge \tau \numberthis
					\end{cases}
			\end{align*}	
			And,
			\begin{align*}
				\Pr & \{\varepsilon_t - \varepsilon_h^o  = \varepsilon_j - \varepsilon_i\} \\
				& = \Pr \{(\varepsilon_j - \varepsilon_{i+1} < \varepsilon_t - \eta' E_h^o \le \varepsilon_j - \varepsilon_i ) \cap (E_h^o < C_2) \} \\
				& \quad + \Pr \{(\varepsilon_j - \varepsilon_{i+1} < \varepsilon_t - \eta' C_2 \le \varepsilon_j - \varepsilon_i ) \cap (E_h^o \ge C_2) \} \\
				& = \begin{cases}
					0 			\qquad \qquad \qquad \quad \;\; \mbox{if } C_2 < \frac{\tau-j+i}{\eta' L}C_1 \\
					\Pr \{\eta' E_h^o \ge \varepsilon_t - \varepsilon_j + \varepsilon_i\}		 \\
					\qquad \qquad \qquad \qquad \mbox{if } \frac{\tau-j+i}{\eta' L}C_1 \le C_2 < \frac{\tau-j+i+1}{\eta' L}C_1 \\
					\Pr \{ \varepsilon_t - \varepsilon_j + \varepsilon_i \le \eta' E_h^o < \varepsilon_t - \varepsilon_i + \varepsilon_{i+1} \} \\
					\qquad \qquad \qquad \qquad \mbox{if } C_2 \ge \frac{\tau-j+i+1}{\eta' L}C_1 \numberthis
					\end{cases} \\
			\end{align*}
			As a result, the transition probability in this case is given by \eqref{eq_p_ji} on the previous page.

		We then examine the stationary distribution $\vect \xi_{FD}$ of the PES energy status, where $\xi_{FD,i}, i\in\{0,1,\dots,L\}$ represents the probability of the residual energy at PES being $\varepsilon_i$. We first define $ \vect{M}_{FD} \triangleq (p_{i,j})$ to denote the $ (L+1)\times(L+1) $ state transition matrix of the MC. By using the similar methods in \cite{krikidis_buffer-aided_2012}, we can easily verify that the MC transition matrix $\vect{M}_{FD}$ is irreducible and row stochastic. Therefore, there must exist a unique stationary distribution $\vect \xi_{FD}$ that satisfies the following equation
		\setcounter{equation}{41}
		\begin{equation} \label{eq_MC_1}
			\vect\xi_{FD} = (\xi_{FD,0}, \xi_{FD,1}, \dots, \xi_{FD,L})^T = (\vect{M}_{FD})^T \vect\xi_{FD}
		\end{equation}
		By solving \eqref{eq_MC_1}, $\vect\xi_{FD}$ can be derived as
		\begin{equation} \label{eq_stationary_distribution_FD}
				\vect{\xi}_{FD} = \left( \left(\vect{M}_{FD}\right)^{T} - \vect{I} + \vect{B}\right)^{-1}\vect{b},
		\end{equation}
		where $ B_{i,j} = 1, \forall i, j $ and $ \vect{b} = (1, 1, \dots, 1)^{T} $.

		We are now ready to derive the probability that the available energy at J mets the energy condition. With the stationary distribution $\vect \xi_{FD}$, we can obtain
		\begin{equation} \label{eq_battery_ready}
			\Pr \left\{\varepsilon[k] \ge E_{th}\right\} = \sum_{i=\tau}^L \xi_{FD,i}
		\end{equation}

	\section{Secrecy Performance Analysis} \label{sec: performance analysis}
		In this section, we characterize the secrecy performance of the proposed AnJ protocol in terms of the secrecy outage probability and the existence of non-zero secrecy capacity. These two probabilistic metrics are widely used in measuring secrecy performance when the eavesdropper's instantaneous CSI is absent. 

		\subsection{Preliminaries} 
			\label{sub:preliminaries}
			The secrecy capacity $C_s$ is defined as the rate difference between the maximum achievable transmission rate of the main channel and that of the wiretap channel \cite{wyner_wire-tap_1975}. Mathematically speaking,
			\begin{equation} \label{eq_cs_defi}
				C_s = \begin{cases}
					C_M - C_W 	& \qquad \text{if}~ \gamma_D > \gamma_E \\
					0			& \qquad \text{if}~ \gamma_D \le \gamma_E
				\end{cases}
			\end{equation}
			where $C_M = \log_2 \; (1+ \gamma_{D})$ is the capacity of the main channel between S and D, and $C_W = \log_2 \; (1+\gamma_{E}) $ is the capacity of the wiretap channel between S and E. In \eqref{eq_cs_defi}, $\gamma_D$ and $\gamma_E$ denote the instantaneous SNRs at D and E, respectively.
			
			Specifically, the secrecy outage probability, i.e., $P_{so}^{AnJ} $, is defined as the probability that the secrecy capacity $C_s$ is less than a target secrecy rate $R_s$\footnote{From \eqref{eq_cs_defi} and \eqref{eq_Pso_defi}, it is clear that secrecy outage must happen if the channel capacity of link S $\to$ D is less than $R_s$, which  motivates our energy condition presented in Section \ref{sub:protocol_description}.}  \cite{barros_secrecy_2006}. Mathematically speaking,
			\begin{equation} \label{eq_Pso_defi}
				P_{so}^{AnJ} = \Pr \{C_s < R_s\}
			\end{equation}

			The existence of non-zero secrecy capacity, i.e., $P_{nzsc}^{AnJ}$, is defined as the probability that the secrecy capacity is greater than zero, i.e.,
			\begin{equation} \label{eq_Pnzsc_defi}
				P_{nzsc}^{AnJ} = \Pr \{C_s > 0\}
			\end{equation}

		\subsection{Secrecy Outage Probability} 
		\label{sub:secrecy_performance_with_perfect_csi}				
			We first derive the signal-to-interference-plus-noise ratio (SINR) and its corresponding PDF and CDF at D and E, respectively.
			From \eqref{eq_y_D}, the SINR at D is given by
			\begin{align} \label{eq_SINR_D}
				\gamma_D = \dfrac{P_S H_{SD}}{ (1-\rho) P_J \sigma_{err}^2 / (N_t-1) + \sigma_{D}^{2}}
			\end{align}
			Since $h_{SD}$ is Rayleigh fading channel, $\gamma_D$ follows an exponential distribution.	

			From \eqref{eq_y_E}, the SINR at E is given by
			\begin{equation} \label{eq_snr_E}
				\gamma_E = \dfrac{P_S|h_{SE}|^{2}}{P_J||(\vect{h}_{JE})^{\dagger} \vect{W}||^{2}/(N_t-1) + \sigma_E^2}
			\end{equation}
			The PDF of $\gamma_E$ depends on $|h_{SE}|^2$ and $||(\vect{h}_{JE})^{\dagger} \vect{W}||^{2}$. To proceed, we first define $X \triangleq P_S|h_{SE}|^2 $. Recall that $h_{SE}$ follows a Rayleigh distribution, we thus have the PDF of $X$ as
			\begin{equation}
				f_X(x) = \frac{1}{P_S \Omega_{SE}}\exp\left(-\frac{x}{P_S\Omega_{SE}}\right)
			\end{equation}
			We also define $Y \triangleq \frac{P_J ||(\vect{h}_{JE})^{\dagger} \vect{W}||^{2}}{N_t - 1}$. Since $||(\vect h_{JE})^\dagger||^2$ is a sum of i.i.d. exponential distributed random variables, and $\vect W$ is a unitary matrix,  $||(\vect{h}_{JE})^{\dagger} \vect{W}||^{2} $ is also a sum of i.i.d. exponential distributed random variables \cite{zhou_secure_2010}. Therefore, $Y$ follows a Gamma distribution $\mathcal{G}(N_t-1, P_J\Omega_{JE}/(N_t-1))$ with the PDF given by
			\begin{align}
				f_Y(y) = \frac{y^{N_t-2} e^{ -\frac{N_t-1}{P_J\Omega_{JE}}y}}{\Gamma(N_t-1)\left(\frac{P_J\Omega_{JE}}{N_t-1}\right)^{N_t-1}}
			\end{align}
			According to \eqref{eq_snr_E}, the expression $\gamma_E = \frac{X}{Y + \sigma_E^2}$ then holds. Therefore, we can obtain the CDF of $\gamma_E$ as
			\begin{align*} \label{eq_CDF_rE}
				F_{\gamma_{E}}(z) & = \Pr\{\gamma_E < z\} = \Pr\left\{\frac{X}{Y + \sigma_E^2} < z\right\} \\
				& = \int_0^\infty \int_0^{zy + z\sigma_E^2} \! f_X(x) f_Y(y) \; \mathrm{d}x \;\mathrm{d}y \\
				& = 1 - e^{-\frac{z \sigma_E^2}{P_S\Omega_{SE}}} \left(\frac{N_t-1}{\varphi z + N_t -1} \right)^{N_t-1} \numberthis
			\end{align*}
			where $\varphi \triangleq \frac{P_J\Omega_{JE}}{P_S\Omega_{SE}}$, and the integral is obtained from \cite[Eq. (3.326.2)]{gradshtein_table_2007}.
			Correspondingly, the PDF of $\gamma_E$ is obtained as
			\begin{align*} \label{eq_PDF_rE}
				f_{\gamma_E}(z) & = \frac{\sigma_E^2 \; e^{-\frac{z\sigma_E^2}{P_S\Omega_{SE}}}}{P_S \Omega_{SE}} \left(\frac{N_t-1}{\varphi z + N_t -1} \right)^{N_t-1} \\
				& +  \varphi e^{-\frac{z\sigma_E^2}{P_S\Omega_{SE}}} \left(\frac{N_t-1}{\varphi z + N_t -1} \right)^{N_t} \numberthis
			\end{align*}

		\begin{proposition} \label{proposition:pso}
		\newcounter{MYtempeqncnt_pso}
			\begin{figure*}[!t]
			\normalsize
			\setcounter{MYtempeqncnt_pso}{\value{equation}}
			\setcounter{equation}{53}
			\begin{align*} \label{eq_pso}
				P_{so}^{AnJ}
				= \begin{cases}
				1 - \left( \frac{\sigma_E^2 }{P_S\Omega_{SE}} \Psi_1(1, \mu_1, \beta_1) +  \Psi_1 (2, \mu_1, \beta_1 ) \right) \varphi^{-1} \exp\left(  -\frac{2^{R_s}-1}{\kappa_1\Omega_{SD}}  \right) \sum_{i=\tau}^L \xi_{FD,i} & \text{if } N_t = 2 \\
				\\ %
				1 - \left(  \frac{\sigma_E^2 }{P_S\Omega_{SE}} \Psi_2 (N_t-1, \mu_1, \beta_1 ) + (N_t-1) \Psi_2 (N_t, \mu_1, \beta_1 ) \right) \beta_1^{N_t-1} \, \exp\left(  -\frac{2^{R_s}-1}{\kappa_1\Omega_{SD}}  \right)\, \sum_{i=\tau}^L \xi_{FD,i} & \text{if } N_t \ge 3 \numberthis
 				\end{cases}
			\end{align*}
			\setcounter{equation}{\value{MYtempeqncnt_pso}}
			\hrulefill
			\vspace*{4pt}
			\end{figure*}
			\setcounter{equation}{54}
			The exact secrecy outage probability for the proposed AnJ protocol is derived as \eqref{eq_pso} at the top of the next page, where $\beta_1 \triangleq (N_t-1)/\varphi$, $\mu_1 \triangleq \frac{2^{R_s}}{\kappa_1\Omega_{SD}} + \frac{\sigma_E^2}{P_S\Omega_{SE}}$, $\kappa_1 \triangleq \frac{P_S}{ (1-\rho) P_J \sigma_{err}^2 / (N_t -1) + \sigma_D^2}$, and
			\begin{equation}
				\Psi_1 (n, \mu, \beta) \triangleq (n-1) \beta^{-1} - (-\mu)^{n-1} e^{\beta\mu} \mathrm{Ei}(-\beta\mu)
			\end{equation}
			\begin{align}
				\Psi_2 (n, \mu, \beta) \triangleq &  \frac{1}{(n-1)! } \sum_{k=1}^{n-1} (k-1)! (-\mu)^{n-k-1} \beta^{-k} \nonumber \\
				& - \frac{(-\mu)^{n-1} }{(n-1)! } e^{\beta\mu} \mathrm{Ei}(-\beta\mu)
			\end{align}
			\begin{IEEEproof}
				See Appendix \ref{appendix: A}.
			\end{IEEEproof}
		\end{proposition}

		\begin{corollary} \label{corollary:pnzsc}
			\newcounter{MYtempeqncnt_positive_Cs}
			\begin{figure*}[!t]
			\normalsize
			\setcounter{MYtempeqncnt_positive_Cs}{\value{equation}}
			\setcounter{equation}{56}
			\begin{align*} \label{eq_p_nzsc}
				P_{nzsc}^{AnJ}
				= \begin{cases}
				\left( \frac{\sigma_E^2 }{P_S\Omega_{SE}} \Psi_1(1, \mu_2, \beta_1+\beta_2) + \Psi_1 (2, \mu_2, \beta_1+\beta_2 ) \right) \varphi^{-1} e^{- \beta_2\mu_2} \sum_{i=\tau}^L \xi_{FD,i} & \text{if } N_t = 2 \\
				\quad + \exp\left(- \frac{2^{R_s}-1}{\kappa_2 \Omega_{SD}} \right) F_{\gamma_E}(\beta_2)  \sum_{i=\tau}^L \xi_{FD,i}  \\
				\\ %
				\left(  \frac{\sigma_E^2 }{P_S\Omega_{SE}} \Psi_2 (N_t-1, \mu_2, \beta_1+\beta_2 ) + (N_t-1) \Psi_2 (N_t, \mu_2, \beta_1+\beta_2 ) \right) \beta_1^{N_t-1} \, e^{- \beta_2\mu_2 } \, \sum_{i=\tau}^L \xi_{FD,i} & \text{if } N_t \ge 3 \\
				\quad + \exp\left(- \frac{2^{R_s}-1}{\kappa_2 \Omega_{SD}} \right) F_{\gamma_E}(\beta_2)  \sum_{i=\tau}^L \xi_{FD,i}  \numberthis
 				\end{cases}
			\end{align*}
			\setcounter{equation}{\value{MYtempeqncnt_positive_Cs}}
			\hrulefill
			\vspace*{4pt}
			\end{figure*}
			\setcounter{equation}{57}
			The probability of non-zero secrecy capacity is derived as \eqref{eq_p_nzsc} at the top of the next page, where $\kappa_2 \triangleq P_S/\sigma_D^2$, $\beta_2 = (2^{R_s}-1)\kappa_1/\kappa_2$, and $\mu_2 \triangleq \frac{1}{\kappa_1\Omega_{SD}} + \frac{\sigma_E^2}{P_S\Omega_{SE}}$.
			\begin{IEEEproof}
				See Appendix \ref{appendix: B}.
			\end{IEEEproof}
		\end{corollary}

		\begin{remark} \label{rmk1}
			For a given source transmit power $P_S$, the probability that the accumulated energy at $J$ is sufficient for jamming decreases with the increase of the jamming power $P_J$. Specifically, the threshold $\tau$ for the last summation term in \eqref{eq_pso} increases with $P_J$. Accordingly, the probability summation $\sum_{i=\tau}^{L}{\xi}_{FD,i}$ decreases with the increase of $P_J$, which has a negative effect on the secrecy outage. On the other hand, according to \eqref{eq_SINR_D} and \eqref{eq_snr_E}, increasing $P_J$ will reduce the SINR at both D and E, but to a lesser extent for $\gamma_D$ compared to $\gamma_E$ as the null space jamming is designed. Consequently, larger $P_J$ can lead to larger instantaneous secrecy capacity $C_s$, which has a positive effect on the secrecy outage. To make a long story short, a higher $P_J$ is associated with lower jamming frequency but higher interference strength. Therefore, we deduce that there would be an optimal $P^*_J$ that minimizes $P_{so}^{\rm AnJ}$. This will be verified by numerical results in Section \ref{sec: simulation}. Unfortunately, due to the complexity of the considered MC model, it is difficult to find a general closed-form solution for $P^*_J$. Nevertheless, for a given network setup, we can readily obtain $P^*_J$ by performing a one-dimensional exhaustive search over the finite range of the discretized energy levels.
		\end{remark}

\section{Continuous Energy Storage Model with Infinite Capacity} 
	\label{sub:infinity capacity energy storage}
	In the proposed AnJ protocol, we employ PES and SES with finite storage capacity at the wireless powered jammer. It is obvious that the system performance can be improved via increasing the capacity of PES and SES: A larger capacity can reduce the energy loss caused by energy overflow, thus the jammer can accumulate more energy for supporting jamming transmission.  On the other hand, one can infer that the rate of the performance improvement actually decreases as the energy storage capacity increases, because energy overflow occurs more rarely as the capacity keeps increasing. Considering the device cost and size, a question then comes up: ``For a given network setup, how much energy storage capacity and the corresponding discretization level are considered as adequate?''  To answer this question, in this subsection, we analyze the upper bound of the system performance with infinite energy storage capacity, i.e., $C_1 \to \infty, C_2 \to \infty$.

	To investigate the long-term behavior of the infinite energy storage, we need to compare the energy consumption $E_{th}$ with $ \mathbb{E}\{\tilde E_h^o\} $, which is the average amount of energy acquired by J in OEH mode. Specifically, $E_{th} < \mathbb{E}\{\tilde E_h^o\}$ means that, on average, the harvested energy in OEH mode can fully meet the required energy consumption at the jammer. In this case, the energy stored in PES steadily accumulates during the communication process, which makes the jammer always meet the energy condition.

	On the other hand, when $E_{th} > \mathbb{E}\{\tilde E_h^o\}$, the harvested energy in OEH mode is, on average, less than the consumed energy. As a result, the energy level at PES stays between zero and $E_{th}$. In this case, the total amount of harvested energy must equal the total amount of energy consumption in the long run. Mathematically, with $q_c$ being the probability of meeting the channel condition, and $q_b$ indicating the probability of activating the energy condition, we have
		\begin{align}
			q_c q_b \mathbb{E}\{\tilde E_h^o\} + (1 - q_c q_b)\mathbb{E}\{E_h^d\} = q_c q_b E_{th} \label{eq_energy_equality_1} \\
			\Rightarrow \quad \quad q_b = \frac{\mathbb{E}\{E_n^d\}}{q_c \left( E_{th} + \mathbb{E}\{E_n^d\} - \mathbb{E}\{\tilde E_h^o\} \right) } \label{eq_energy_equality_2}
		\end{align}

	With the CDF of $H_{SJ}^d$ in \eqref{eq_CDF_HSJd}, we can calculate $\mathbb{E}\{E_n^d\}$ as
	\begin{align*} \label{eq_Ehd_expectation}
		\mathbb{E}\{E_n^d\}  & = \eta P_S \mathbb{E}\{H_{SJ}^d\} \\
					& = \eta P_S \int_0^\infty \! x F'_{H_{SJ}^d}(x) \; \mathrm{d}x = \eta P_S N_J\Omega_{SJ} \numberthis
	\end{align*}

	Similarly, with the CDF of $H_{SJ}^o$ in \eqref{eq_F_HSJ_o}, we have
	\begin{align*} \label{eq_Eho_expectation}
		\mathbb{E}\{\tilde E_h^o\}
		& = \eta \eta' P_S \mathbb{E}\{H_{SJ}^o\} = \eta \eta' P_S N_r\Omega_{SJ} \numberthis    \\
	\end{align*}

	When combining \eqref{eq_Ehd_expectation} and \eqref{eq_Eho_expectation} with \eqref{eq_energy_equality_2}, we can obtain $q_b$ as
	\begin{align} \label{eq_q}
		q_b = \frac{\eta P_S N_J\Omega_{SJ}}{ q_c \left( E_{th}
		+ \eta P_S N_J\Omega_{SJ}
		- \eta \eta' P_S N_r\Omega_{SJ}
		\right) }
	\end{align}
	Note that $q_c$ is already given in \eqref{eq_channel_ready}.

	\begin{corollary} \label{corollary:pso_HD}
			The closed-form expression of the secrecy outage probability for a cooperative jammer with an infinite capacity energy storage can be obtained by replacing $ \sum_{i=\tau}^L \xi_{FD,i}$ in \eqref{eq_pso} with $q_b$.
	\end{corollary}



		 		
	\section{Cooperative Jamming by a Wireless-Powered Half-Duplex Jammer} \label{sec:HD_jammer}
		In this subsection, we consider an alternative cooperative jamming protocol with a wireless-powered HD jammer J' to provide a benchmark for evaluating the performance of the proposed AnJ protocol.

		In order to compare our proposed FD jammer and this HD jammer in a fair manner, we assume that J' is equipped with the same number of antennas and rectifiers as J, i.e., $N_J$ RF antennas and $N_J$ rectifiers. All antennas and rectifiers are connected in a non-permanent manner. Due to the HD mode, J' requires only one energy storage. We let the capacity of the energy storage at J' be $C_1$, same as that of the PES at J. It is noteworthy that the analytical approach in \cite{liu_secure_2015} for a HD jammer is not applicable to J', because our energy storage has a finite capacity, while the battery capacity in \cite{liu_secure_2015} was infinite.
		Similar to J, J' also operates in two modes, i.e., the energy harvesting (EH) mode and the cooperative jamming (CJ) mode. In EH mode, J' performs exactly like J and harvests the same amount of energy $E_h^d$. All acquired energy is saved in its single energy storage. In CJ mode, on the contrary, J' uses all $N_J$ antennas to transmit jamming signals, and therefore, does not acquire any energy. Assuming that imperfect channel estimation also occurs at J', the corresponding SINR at D and E are given by
		\setcounter{equation}{69}
		\begin{align} \label{eq_SINR_D_HD}
				\gamma_D' = \dfrac{P_S H_{SD}}{ (1-\rho) P_J \sigma_{err}^2 / (N_J-1) + \sigma_{D}^{2}}
		\end{align}
		and
		\begin{equation} \label{eq_snr_E_HD}
				\gamma_E' = \dfrac{P_S|h_{SE}|^{2}}{P_J||(\vect{h}_{JE})^{\dagger} \vect{W}||^{2}/(N_J-1) + \sigma_E^2}
		\end{equation}
		Note that \eqref{eq_SINR_D_HD} and \eqref{eq_snr_E_HD} differ from \eqref{eq_SINR_D} and \eqref{eq_snr_E} only in the number of transmitting antennas, i.e., $N_r$ is replaced with $N_J$.

		\begin{remark} \label{rmk:2}
			Owing to the FD operation mode, it is evident that J harvests more energy than J'. Yet, how much additional energy that J can acquire mainly depends on how the antennas at J are assigned for energy harvesting and jamming transmission.  The impact of antenna allocation on secrecy performance of the proposed AnJ protocol will be shown in numerical results.
		\end{remark}

		\subsection{Markov Chain for HD Jammer} 
			\label{sub:battery_discretization_and_markov_chain}
			We discretize the battery at J' in the way as described in Section \ref{sec_sub: battery discretization} (i.e., the same discretization as J). Different from the MC presented in Section \ref{subsec: markov chain}, all state transitions at J' without a decrease in energy levels refer to the EH mode, because no energy is harvested by J' in the CJ mode. The state transition probability of J', $p'_{i,j}$, is characterized in the following six cases. Due to space scarcity, we have skipped the full details. 	
		\subsubsection{The battery remains empty ($S_0 \to S_0$)}
			\begin{align}
				p'_{0,0}  = \Pr \left\{ \varepsilon_h^d = 0 \right\} = F_{H_{SJ}^d} \left(\frac{1}{\eta P_S L/C_1}\right)
			\end{align}
		\subsubsection{The battery remains full ($S_L \to S_L$)}
			\begin{align}
				p'_{L,L}  = \Pr \left\{C_{SD} < R_s \right\} = 1 - q_c
			\end{align}
		\subsubsection{The non-empty and non-full PES remains unchangeable ($S_i \to S_i: 0 < i < L$)}
			\begin{align*}
				p'_{i,i}
				& = \Pr \{\varepsilon_i < \varepsilon_t\} \Pr \{\varepsilon_h^d = 0\} \\
						& \quad + \Pr \{\varepsilon_i \ge \varepsilon_t\} \Pr\{C_{SD} < R_s \} \Pr \{\varepsilon_h^d = 0\} \\
						& =  \begin{cases}
							F_{H_{SJ}^d}\left(\frac{1}{\eta P_S L/C_1}\right)  & \mbox{if } i < \tau \numberthis \\
							(1 - q_c) F_{H_{SJ}^d}\left(\frac{1}{\eta P_S L/C_1}\right)  & \mbox{if } i \ge \tau \\
						\end{cases}
			\end{align*}
		\subsubsection{PES is partially charged  ($S_i \to S_j: 0 \le i < j < L$)} 
			\begin{align*}
				p'_{i,j} = & \Pr \{\varepsilon_i < \varepsilon_t\} \Pr \{\varepsilon_h^d = \varepsilon_j - \varepsilon_i\} \\
						& + \Pr \{\varepsilon_i \ge \varepsilon_t\} \Pr\{C_{SD} < R_s \} \Pr \{\varepsilon_h^d = \varepsilon_j - \varepsilon_i\} \\
						= & \begin{cases}
							F_{H_{SJ}^d} \left(\frac{j-i+1}{\eta P_S L/C_1}\right) - F_{H_{SJ}^d} \left(\frac{j-i}{\eta P_S L/C_1}\right)  & \mbox{if } i < \tau \numberthis \\
							\left(F_{H_{SJ}^d}\left(\frac{j-i+1}{\eta P_S L/C_1}\right) - F_{H_{SJ}^d}\left(\frac{j-i}{\eta P_S L/C_1}\right)\right) & \mbox{if } i \ge \tau  \\
							\quad \times (1 - q_c)
						\end{cases}
			\end{align*}
		\subsubsection{The battery is fully charged  ($S_i \to S_L: 0 \le i < L$)} 
			\begin{align*}
				p'_{i,L}
				& = \Pr \{\varepsilon_i < \varepsilon_t\} \Pr \{\varepsilon_h^d \ge \varepsilon_L - \varepsilon_i\} \\
						& \quad + \Pr \{\varepsilon_i \ge \varepsilon_t\} \Pr\{C_{SD} < R_s \} \Pr \{\varepsilon_h^d \ge \varepsilon_L - \varepsilon_i\} \\
						& = \begin{cases}
							1 - F_{H_{SJ}^d}\left( \frac{L-i}{\eta P_{S} L/C_1}\right) & \mbox{if } i < \tau \numberthis \\
							(1 - q_c) \left(1 - F_{H_{SJ}^d}\left( \frac{L-i}{\eta P_{S} L/C_1}\right)\right) & \mbox{if } i \ge \tau \\
						\end{cases}
			\end{align*}	
		\subsubsection{The battery is discharged  ($S_j \to S_i: 0 \le i < j \le L$)} 
			\begin{align*}
				p'_{j,i}
				& = \Pr \{\varepsilon_j \ge \varepsilon_t\} \Pr\{C_{SD} \ge R_s \} \Pr \{\varepsilon_t = \varepsilon_j - \varepsilon_i\} \\
				& = \begin{cases}
					q_c & \text{if } i = j - \tau \\
					0  & \mbox{otherwise} \numberthis
					\end{cases}
			\end{align*}

		Based on the above expressions for $p'_{i,j}$, we define the transition matrix of the above MC as $ \vect{M}_{HD} \triangleq (p'_{i,j})$. Similar as \eqref{eq_stationary_distribution_FD}, the stationary distribution of the battery at J' is given by
		\begin{equation} \label{eq_stationary_distribution_HD}
			\vect{\xi}_{HD} = (\vect{M}_{HD}^{T} - \vect{I} + \vect{B})^{-1}\vect{b}.
		\end{equation} where the $i$-th entry, $\xi_{HD,i}$, denotes the probability of the residual energy in the battery of J' being $\varepsilon_i$. As a result, the probability for the energy condition being met at J' is
		\begin{equation} \label{eq_battery_ready_HD}
			\Pr \left\{\varepsilon[k] \ge E_{th}\right\} = \sum_{i=\tau}^L \xi_{HD,i}
		\end{equation}

		\subsection{Secrecy Performance for HD Jammer} \label{sub:secrecy_performance_HD}
		In this subsection, we derive the secrecy performance for the cooperative jamming scheme with J'. Based on Proposition \ref{proposition:pso} and Corollary \ref{corollary:pnzsc}, we can obtain the following corollaries.
		\begin{corollary} \label{corollary:pso_HD}
			The closed-form expression of the secrecy outage probability for cooperative jamming from J' can be obtained by replacing $N_t$ and $ \xi_{FD,i}$ in \eqref{eq_pso} with $N_J$ and $ \xi_{HD,i} $, respectively.
		\end{corollary}
		
		\begin{corollary} \label{corollary:pnzsc_HD}
			The closed-form expression of the probability of non-zero secrecy capacity for cooperative jamming from J' can be obtained by replacing $N_t$ in \eqref{eq_p_nzsc} with $N_J$, and replacing $ \xi_{FD,i} $ with $ \xi_{HD,i} $.
		\end{corollary}

	\section{Numerical Results} \label{sec: simulation}
		In this section, We provide numerical results based on the analytical expressions developed in the previous sections, and investigate the impact of key system parameters on the performance. In line with \cite{liu_secure_2015}, the simulation is carried out on of a linear topology where nodes S, J, E, and D are placed in order along a horizontal line; the distances are set to $d_{SJ} = 5 $ m, $d_{SE} = 20 $ m and $d_{SD} = 30 $ m. Throughout this section,  unless otherwise stated, we set the path loss exponent $\alpha = 3$, the fading channel variances $\Omega_{ij} = 1/(1+d_{ij}^{\alpha}) $, the noise power $\sigma_D^2 = -80 $ dBm, the target secrecy rate $R_s = 1$, the Rician factor $K = 5$ dB, the channel estimation factor $\rho = 1$, and the number of antennas at the jammer $N_J = 8$ (i.e., $N_r = N_t = 4$). For parameters regarding the energy storage, we set the energy conversion efficiency $\eta = 0.5$, the energy transfer efficiency $\eta' = 0.9$, the PES capacity $C_1 = 0.02$, the SES capacity $C_2 = 0.01$, the discretization level $L = 100$, and the constant circuitry power $P_c = 0.1\times 10^{-3}$ watt\footnote{We note that typical values for practical parameters used in EH systems depend on both the system application and specific technology used for implementation of RF energy harvesting circuits.}.

		\begin{figure}[!t]
			\centering
			\includegraphics[width=21pc]{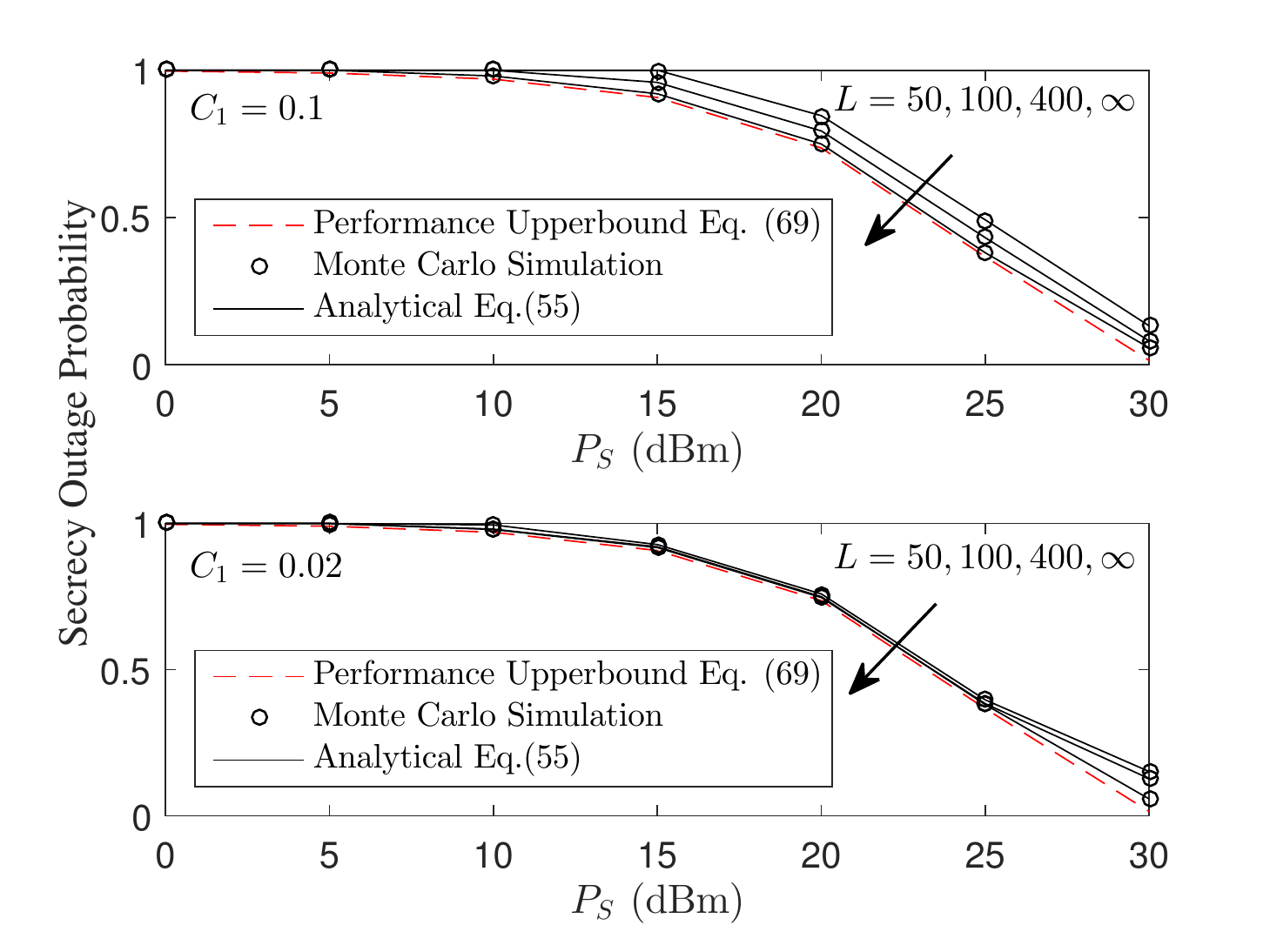}
			\caption{Secrecy outage probability with various $C_1$ and $L$ versus $P_S$. $ P_J = 10 \text{ dBm}$}
			\label{fig_pso_C1_L}
		\end{figure}
	\subsection{The Validation of Energy Discretization Model} 
		\label{sub:the_validation_of_battery_model}
		In this subsection, we examine the accuracy of the energy discretization model (referred to as EDM hereafter, for notation simplicity) presented in Section \ref{sec: markov chain}. Fig. \ref{fig_pso_C1_L} shows the secrecy outage probability with different PES capacity values and discretization levels. The performance of the continuous energy storage with infinite capacity is also plotted to serve as an upper bound. In the case of $C_1 = 0.1$, it can be seen from the figure that the performance of EDM approaches the upper bound as $L$ increases. Specifically, when $L = 400$, the performance of EDM coincides with the upper bound. This is because a larger $L$ results in a smaller quantization step size, i.e., $C_1/L$, for a given PES capacity $C_1$. As a result, the energy loss caused by the discretization process reduces. On the other hand, when $C_1 = 0.02$, it is observed that the performance of EDM converges to the upper bound much more rapidly than the case of $C_1 = 0.1$. In particular, even a small discretization level of $L = 50$ suffices the close match. This is because  given a small $C_1$, a small value of $L$ is adequate to provide the same discretization granularity. This observation allows the system designer to reduce computation via choosing a small $L$, when the energy storage capacity is low. Besides, when $P_S$ exceeds $25$ dBm, the performance of EDM deviates from the upper bound, which indicates that energy overflow occurs frequently, and the selected storage capacity should be enlarged.
		
		\begin{figure}[!t]
			\centering
			\includegraphics[width=21pc]{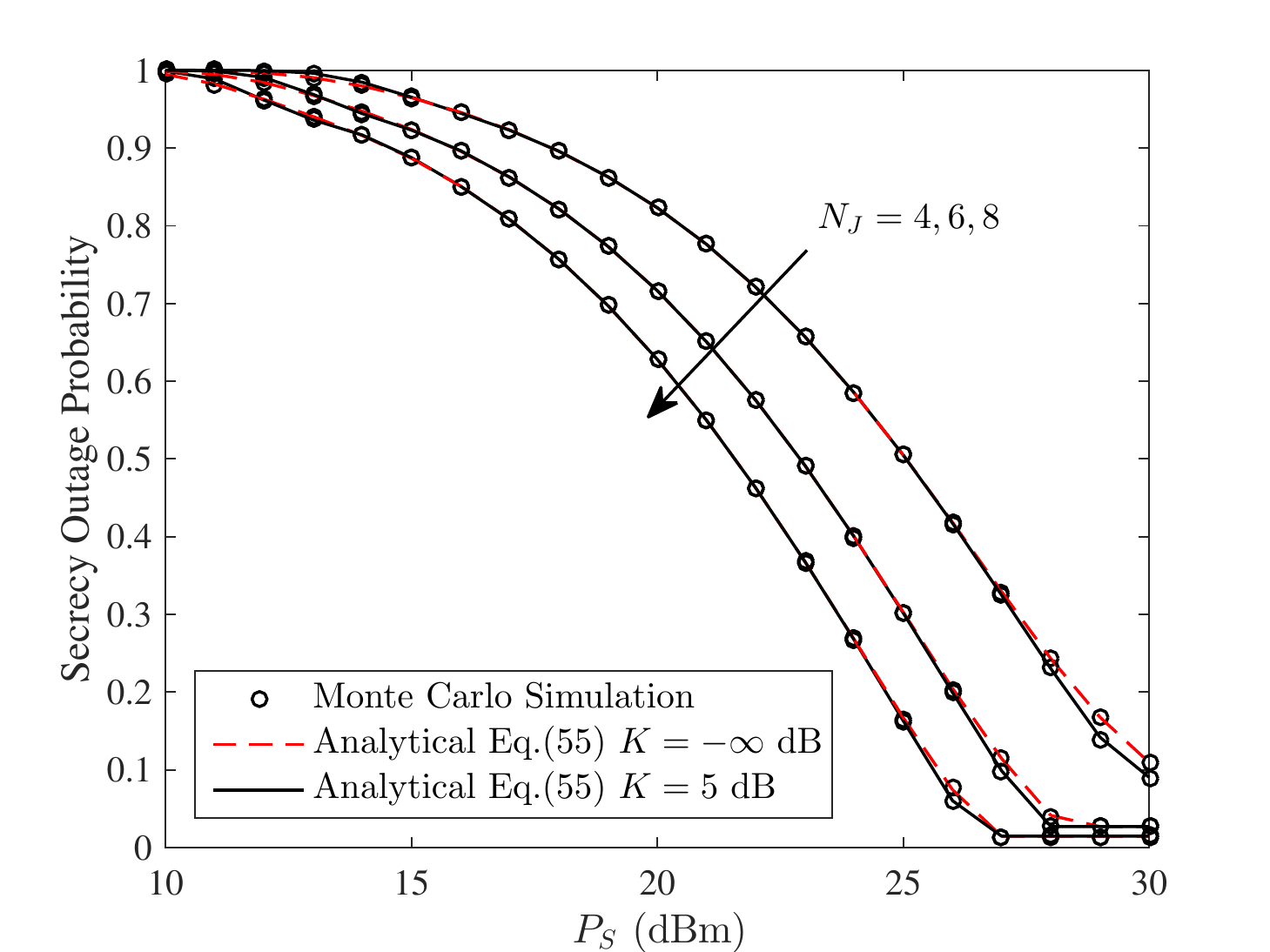}
			\caption{Secrecy outage probability with various $N_J$ and $K$ versus $P_S$. $P_J = 0 \text{ dBm}$}
			\label{fig_pos_Nj_K}
		\end{figure}
		\begin{figure}[!t]
			\centering
			\includegraphics[width=21pc]{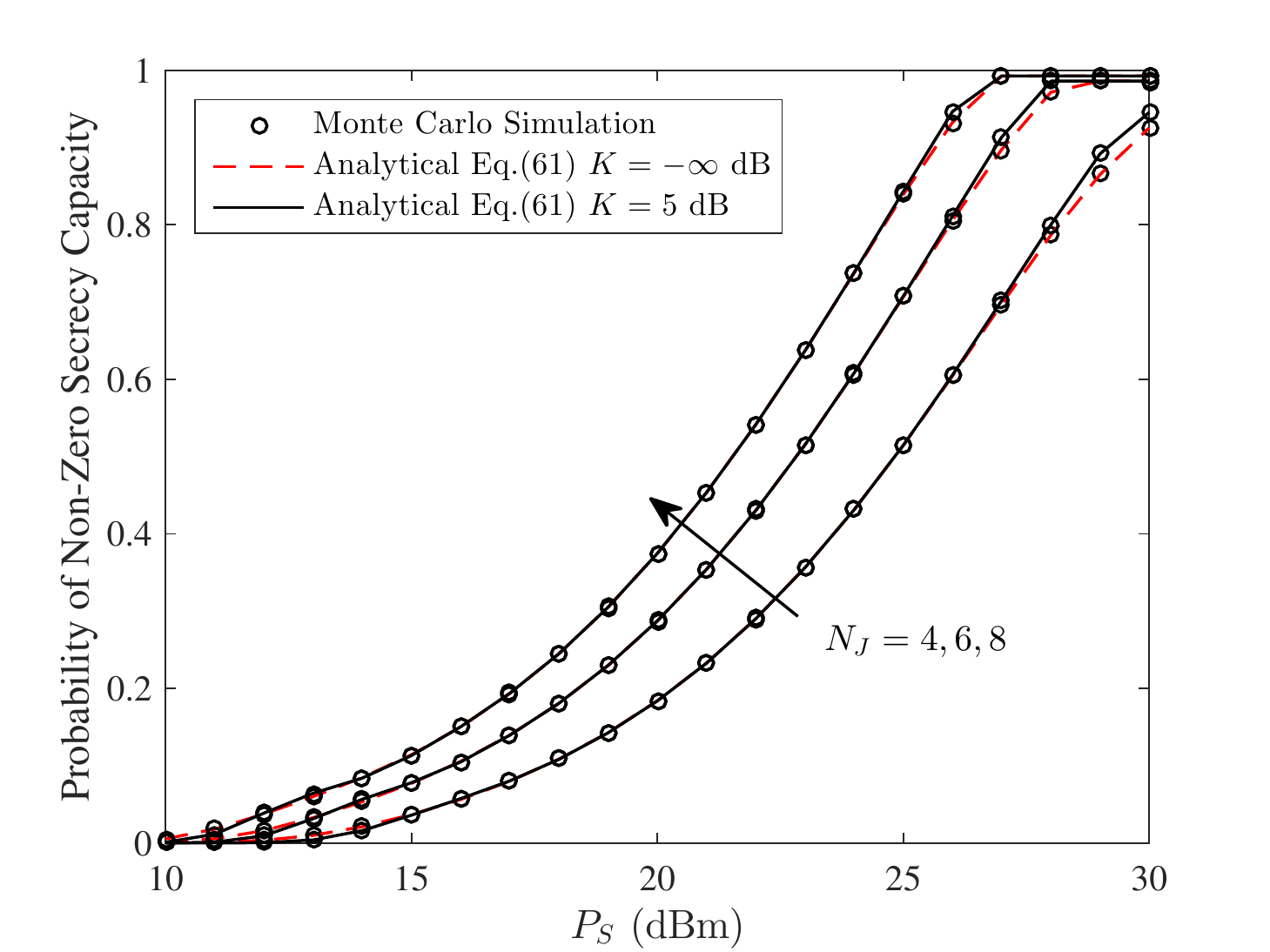}
			\caption{The existence of non-zero secrecy capacity with various $N_J$ and $K$ versus $P_S$. $P_J = 0 \text{ dBm}$}
			\label{fig_pnzsc_Nj_K}
		\end{figure}


	\subsection{The Effect of the Number of Jammer Antennas and Rician Factor} 
		\label{sub:the_validation_of_secrecy_metrics}
		In this subsection, we investigate the effects of the number of antennas at the jammer (i.e., $N_J$) and the Rician factor (i.e., $K$) on the secrecy performance. In Fig. \ref{fig_pos_Nj_K}, the solid lines are for $K = 5$ dB (i.e., Rician fading), whereas the dashed lines are for $K = -\infty$ dB (i.e., Rayleigh fading). The performance differences between these two are surprisingly minor, which indicates that the strength of the LoS path between S and J has limited impact on the system performance. On the contrary, the effect of $N_J$ is remarkable: As $N_J$ increases from $4$ to $8$, the secrecy outage decreases significantly. In addition, the increase of $P_S$ also improves the performance notably. The positive association of $N_J$ and $P_S$ with system performance is because greater $N_J$ and/or $P_S$ can increase the amount of harvested energy at the jammer, and therefore can support more frequent jamming. The finding suggests that increasing the number of antennas at the jammer and/or increasing the transmitting power at the source are two effective manners for secrecy improvement. Monte Carlo simulation results are also provided in Fig. \ref{fig_pos_Nj_K} to validate the closed-form expressions in Eq. \eqref{eq_pso}. In addition, similar positive effects of $N_J$ and $K$ on the probability of non-zero secrecy capacity can be observed in Fig. \ref{fig_pnzsc_Nj_K}. Monte Carlo simulation results presented in Fig. \ref{fig_pnzsc_Nj_K} are in line with the closed-form expressions in Eq. \eqref{eq_p_nzsc}.

	\subsection{The Effect of the Jamming Power} \label{sub:the_effect_of_the_jamming_power}
\begin{figure}[!t]
			\centering
			\includegraphics[width=21pc]{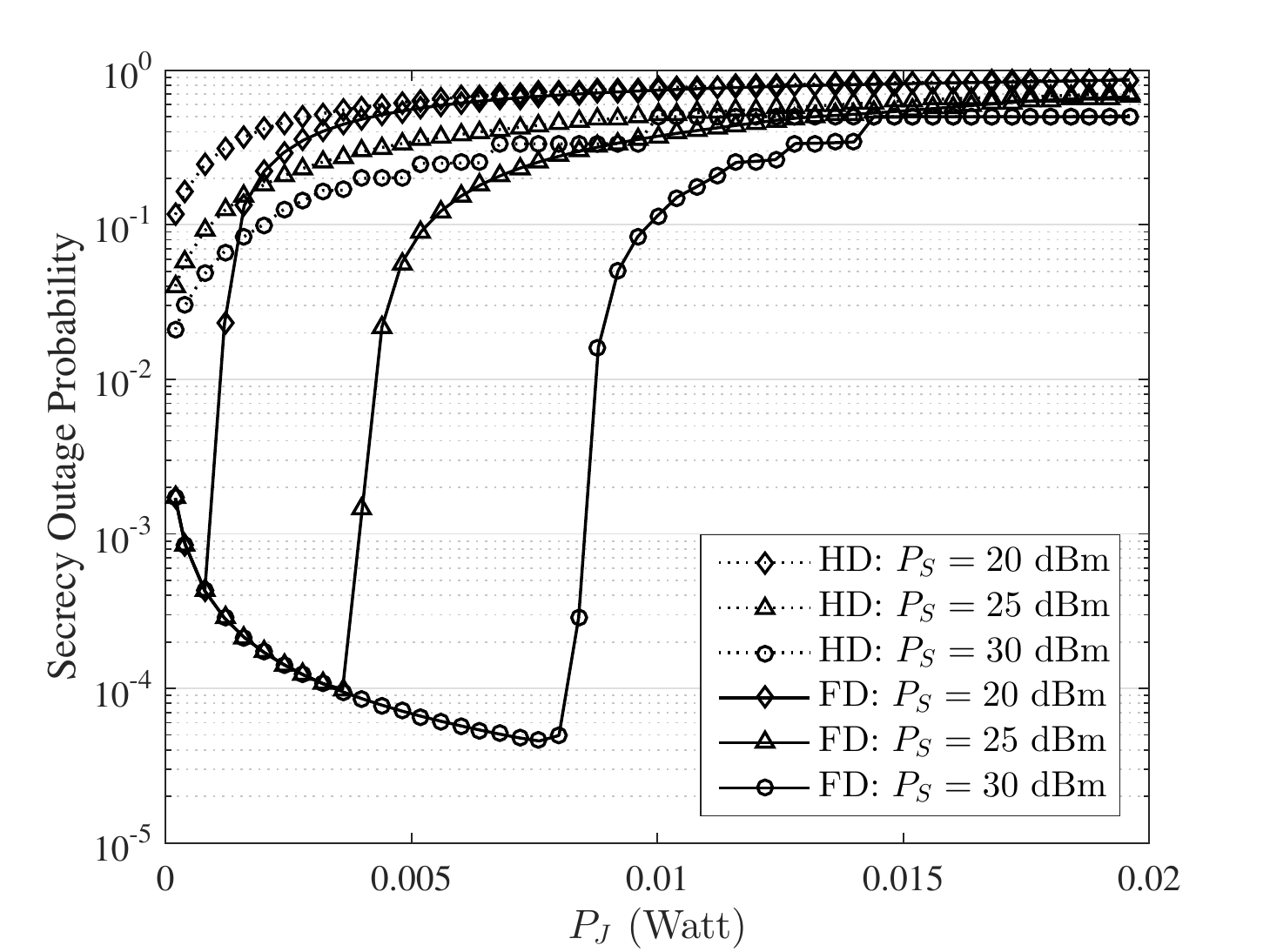}
			\caption{Secrecy outage probability with various $P_S$ versus jamming power $P_J$.}
			\label{fig_opt_pj}
		\end{figure}

		Fig. \ref{fig_opt_pj} shows the association between the secrecy outage probability and the jamming power $P_J$. The source transmitting power is chosen from $P_S = [20, 25, 30]$ dBm. Overall, it can be seen that a distinct optimum jamming power $P^*_J$ with the minimum secrecy outage probability, exists in all considered scenarios. The existence of $P^*_J$ is because, in short, a higher $P_J$ is associated with lower jamming frequency but higher interference strength. This finding validates our deduction in Remark \ref{rmk1}. It also implies that in a scenario with multiple jammers, the jamming power at each jammer should be individually optimized. In addition, as expected, the optimum jamming power of the proposed FD scheme is notably higher than that of HD.
		
	\subsection{The Performance Comparison Between FD Jamming and HD Jamming}
	\label{sub:the_comparison_between_FD_and_HD}
\begin{figure}[!t]
			\centering
			\includegraphics[width=21pc]{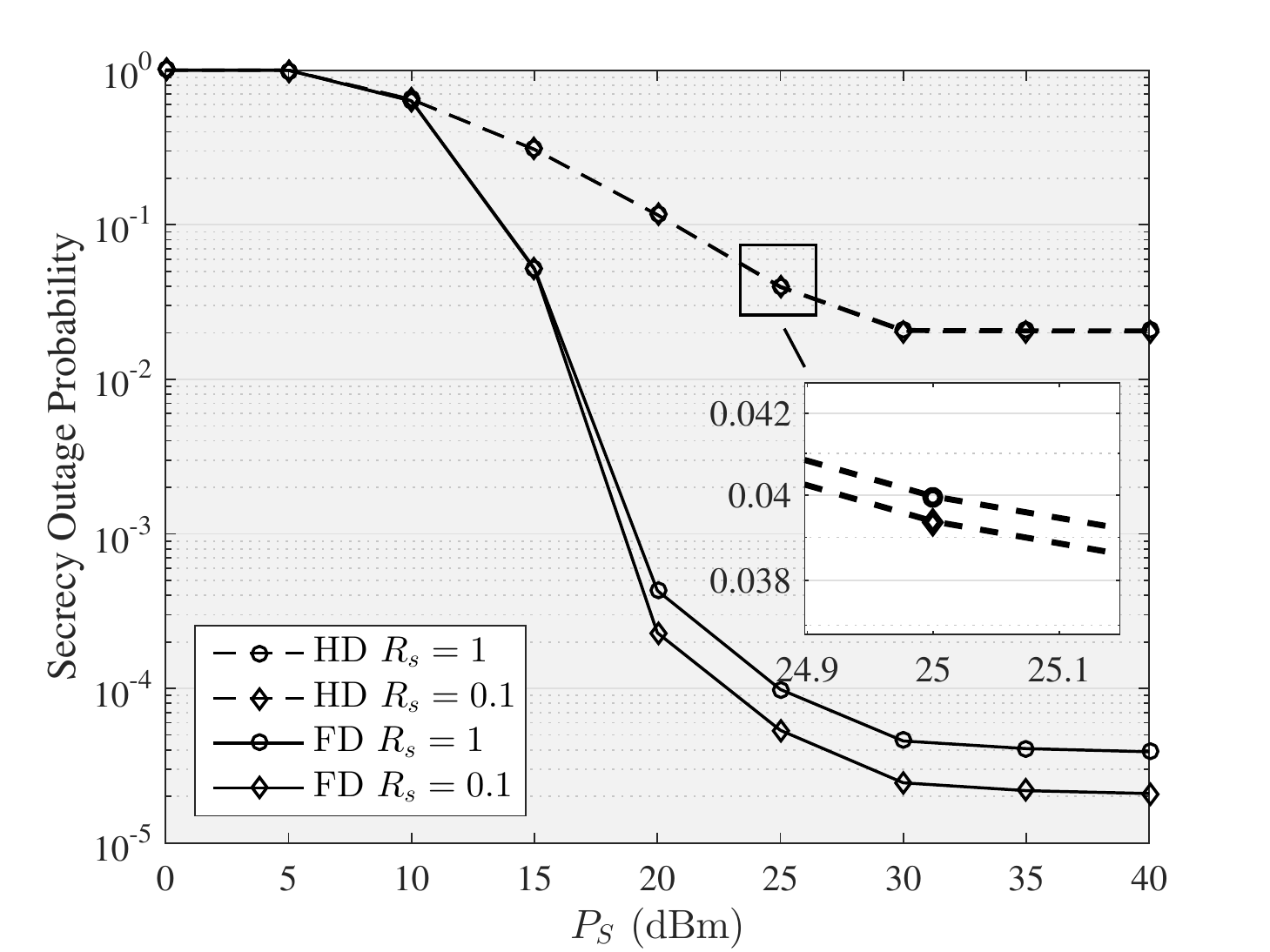}
			\caption{Comparison of the optimum secrecy outage probability of FD and HD, $P_J = P_J^*$ }
			\label{fig_pos_FD_HD}
		\end{figure}

		In this subsection, we compare the optimum secrecy performance between the FD and the HD scheme. Fig. \ref{fig_pos_FD_HD} illustrates the secrecy outage probability for the two schemes with various secrecy rate $R_s$. The jamming power for both FD and HD is chosen to be the corresponding optimum value, i.e., $P_J = P^*_J$. It is clear from the figure that the proposed FD scheme achieves significantly lower secrecy outage than the HD scheme over the entire range of $P_S$. Specifically, when $R_s$ reduces from $1$ to $0.1$, the reduction in secrecy outage for the FD scheme is more notable than that for the HD scheme, suggesting that reducing $R_s$ as a method to mitigate the outage is more effective in the FD scheme than in the HD scheme. Moreover, the performance gap between the two schemes can be further enlarged by rearranging the antenna allocation at the FD jammer, which will be discussed in the next subsection.

	\subsection{The Effect of Antenna Allocation at the Jammer} 
	\label{sub:the_effect_of_antenna_splitting_ratio}
\begin{figure}[!t]
			\centering
			\includegraphics[width=21pc]{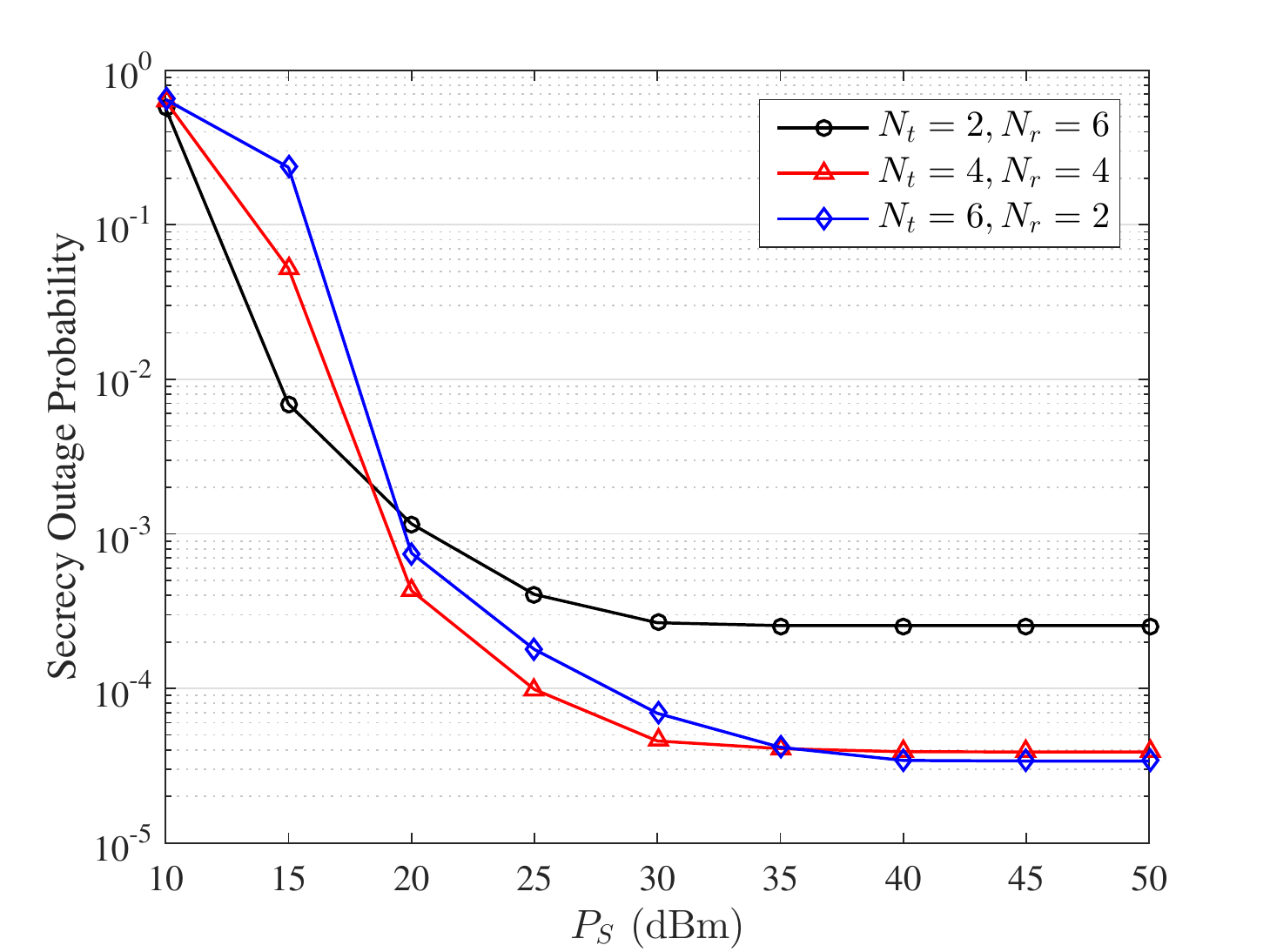} 
			\caption{Secrecy outage probability with various antenna allocation at the jammer, $P_J = P^*_J$.}
			\label{fig_split_antenna}
		\end{figure}

		In this subsection, we investigate the impact of antenna allocation at the jammer on system performance. Fig. \ref{fig_split_antenna} shows the secrecy outage probability of the proposed protocol with different transmitting/receiving antenna allocations. When $P_S$ increases from $10$ dBm to $15$ dBm, the allocation of $N_t = 2, N_r = 6$ achieves the smallest secrecy outage; after $15$ dBm, equal allocation of $N_t = 4, N_r = 4$ overtakes until $P_S$ increases to $35$ dBm. In the high transmitting power regime, the allocation of $N_t = 6, N_r = 2$ finally catches up. The finding suggests that in the cases that the source is sending with low transmitting power, more antennas should be used for energy harvesting, whereas in the paradigm where the source is sending with high transmitting power, more antennas should be used for cooperative jamming, as fewer antennas are required to receive sufficient energy.

		\begin{figure}[!t]
			\centering
			\includegraphics[width=21pc]{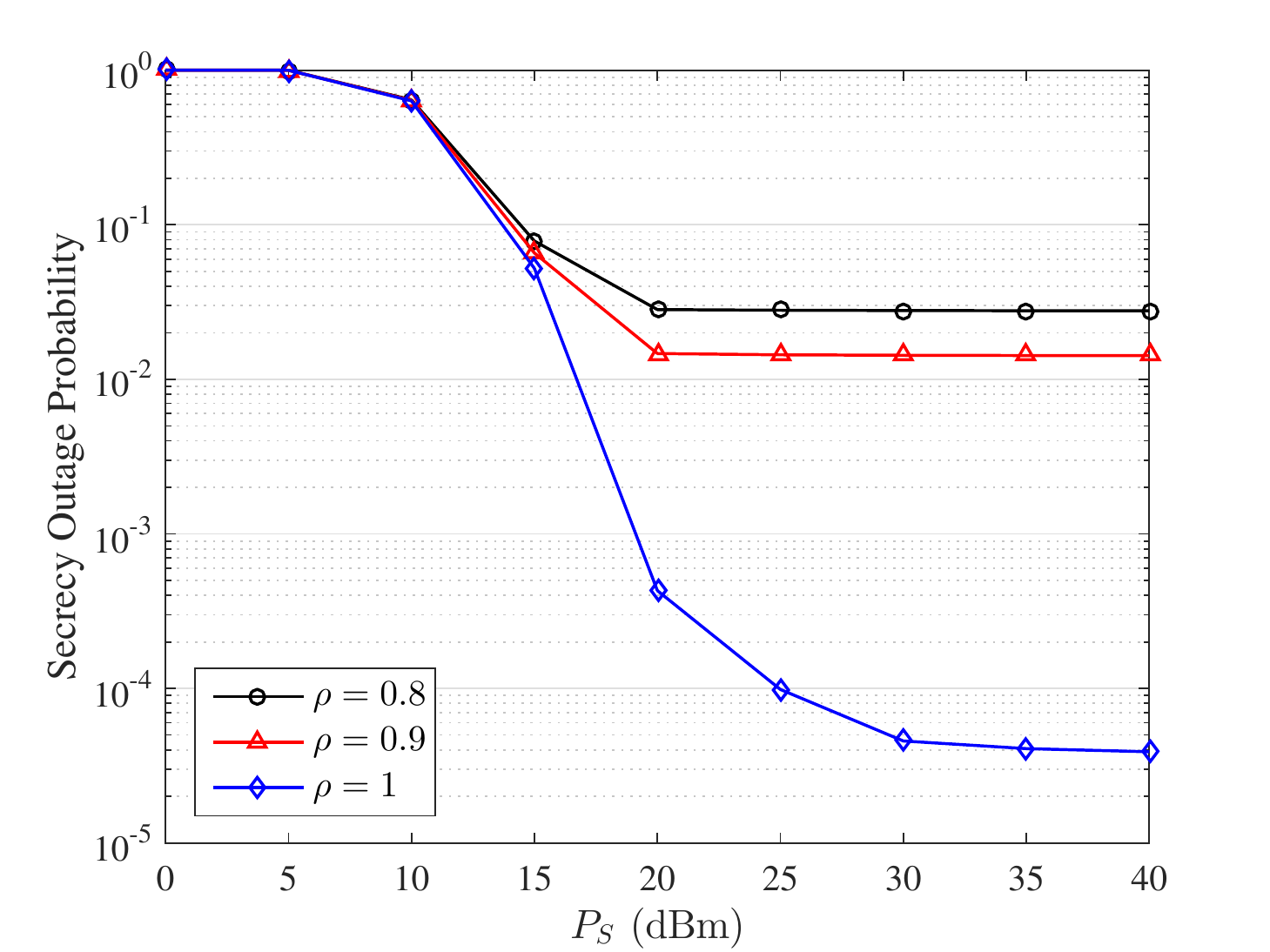}
			\caption{Secrecy outage probability with various $\rho$ versus $P_S$, $P_J = P^*_J$. }
			\label{fig_pos_rho}
		\end{figure}	

	\subsection{The Effect of Channel Estimation Error} 
	\label{sub:the_effect_of_channel_estimation_error}
		In this subsection, we investigate the impact of $\rho$ on the secrecy outage probability of the proposed protocol. From Fig. \ref{fig_pos_rho}, as expected, the CSI mismatch indeed results in performance loss. Specifically, the performance loss is dramatic when slightly reducing $\rho$ from $1$ to $0.9$. The finding indicates that the system performance in practice can be severely degraded by imperfect CSI. Therefore, developing advanced CSI estimation technique dedicatedly for wireless-powered communication network is critical for physical layer security.

	\section{Conclusion} \label{sec: conclusion}
		This paper investigates and discusses 1) the dynamic charging/discharging behavior of the finite capacity energy storage at the jammer, and 2) the secrecy outage probability and the existence of the non-zero secrecy capacity of the proposed AnJ protocol. For the former one, we applied energy discretization model and Markov Chain to derive its stationary distribution over the long term. For the latter one, we took into account the imperfect CSI at the jammer. Additionally, we investigated an infinite capacity energy storage to reveal the performance upper bound. We also derived the secrecy metrics of a wireless-powered half-duplex jammer to serves as a performance benchmark. Numerical results demonstrate that our proposed protocol can provide not only a superior performance over the conventional half-duplex schemes, but also a satisfactory performance close to the upper bound when the energy storage is sufficiently subdivided.
	
	
	
\appendices
	\section{Proof of Proposition 1} \label{appendix: A}
		From the definition of secrecy outage probability given in \eqref{eq_Pso_defi}, by applying the total probability theorem, $P_{so}^{AnJ}$ can be expressed as
		\begin{align*}
			P_{so}^{AnJ} = & \underbrace{ \Pr \{ C_s < R_s | \Phi = \Phi_d \} \Pr \{\Phi = \Phi_d \} }_{\ell_1}  \\
			& + \underbrace{ \Pr \{ C_s < R_s | \Phi = \Phi_o \} \Pr \{\Phi = \Phi_o \} }_{\ell_2}  \numberthis
		\end{align*}

		First, we evaluate the secrecy outage in DEH mode. Recall that no secret data is transmitted in DEH blocks, $\gamma_D$ and $\gamma_E$ hence both equal zero, and further, $C_s $ equals $0$. As $R_s$ is positive, it can be inferred that $\Pr \{ C_s < R_s | \Phi = \Phi_d \} = 1$. Therefore we have $\ell_1 = \Pr\{\Phi = \Phi_d\}$.
		Invoking the independence between the channel condition and the energy condition, and also combining \eqref{eq_channel_ready} and \eqref{eq_battery_ready}, we can obtain
		\begin{align*} \label{eq_ell_1}
			\ell_1 	& = 1 - \Pr \{ (C_{SD} \ge R_s) \cap (\varepsilon[k] \ge E_{th}) \} \\
			& = 1 - \left(1 - F_{H_{SD}} \left( \frac{2^{R_s}-1}{P_S/\sigma_D^2} \right)\right) \sum_{i=\tau}^L \xi_{FD,i} \numberthis
		\end{align*}

		Next, we evaluate the secrecy outage in OEH mode. Considering that $C_s$ and $C_{SD}$ are not independent with each other, but both are independent with the energy random variables, we recast $\ell_2$ as,
		\begin{align*} \label{eq_ell_2}
			\ell_2
				& = \Pr \{ (C_s < R_s) \cap (C_{SD} \ge R_s) \cap (\varepsilon[k] \ge E_{th}) \} \\
				& = \underbrace{\Pr \{ (C_s < R_s) \cap (C_{SD} \ge R_s)\} }_{\ell_{A}} \sum_{i=\tau}^L \xi_{FD,i} \numberthis
		\end{align*}
		Substituting \eqref{eq_channel_condition}, \eqref{eq_Pso_defi} and \eqref{eq_SINR_D} into \eqref{eq_ell_2}, and performing basic mathematical manipulations, we obtain
		\begin{align*} \label{eq_ell_A}
			& \ell_A \\
			& = \Pr \left\{\left( H_{SD} < \frac{ (1+\gamma_E) 2^{R_s} -1 }{\kappa_1} \right) \bigcap \left( H_{SD} \ge \frac{2^{R_s}-1}{\kappa_2} \right) \right\} \\
			& = \Pr \left\{ \frac{2^{R_s}-1}{\kappa_2} \le H_{SD} < \frac{ (1+\gamma_E) 2^{R_s} -1 }{\kappa_1}  \right\}  \\
			& = \displaystyle \int_0^\infty \int_{\frac{2^{R_s}-1}{\kappa_2}}^{\frac{ (1+\gamma_E) 2^{R_s} -1 }{\kappa_1}} f_{H_{SD}}(H_{SD}) \! f_{\gamma_E}(\gamma_E) \mathrm{d}H_{SD} \; \mathrm{d} \gamma_E \numberthis
		\end{align*}
		where $f_{H_{SD}}(\cdot)$ represents the PDF of $H_{SD}$ and
		\begin{equation}
			\kappa_1 \triangleq \frac{P_S}{ (1-\rho) P_J \sigma_{err}^2 / (N_t -1) + \sigma_D^2},\; \kappa_2 \triangleq \frac{P_S}{\sigma_D^2}
		\end{equation}
		Substituting \eqref{eq_CDF_HSD} together with \eqref{eq_PDF_rE} into \eqref{eq_ell_A}, and applying \cite[Eq. (3.352.4) and (3.353.2)]{gradshtein_table_2007} to solve the resultant integrals, we derive $\ell_A$ as
		\begin{align*}
		\ell_A = \begin{cases}
					\exp \left( -\frac{2^{R_s}-1}{\kappa_2\Omega_{SD}} \right) & \text{if } N_t = 2 \\
					\quad - \frac{  \sigma_E^2 \beta_1 }{P_S\Omega_{SE}} \exp\left(-\frac{2^{R_s}-1}{\kappa_1\Omega_{SD}}\right) \Psi_1(1, \mu, \frac{N_t-1}{\varphi})  \\
					\quad - \varphi \beta_1^2 \exp\left(-\frac{2^{R_s}-1}{\kappa_1\Omega_{SD}}\right) \Psi_1 (2, \mu, \frac{N_t-1}{\varphi} ) \\
					\exp \left( -\frac{2^{R_s}-1}{\kappa_2\Omega_{SD}} \right)   & \text{if } N_t \ge 3 \numberthis \\
					\quad - \frac{  \sigma_E^2 \beta_1^{N_t-1} }{P_S\Omega_{SE}} \exp\left(-\frac{2^{R_s}-1}{\kappa_1\Omega_{SD}}\right) \\
					\quad \times  \Psi_2 (N_t-1, \mu, \beta_1 ) \\
					\quad- \varphi  \beta_1^{N_t}  \exp\left(-\frac{2^{R_s}-1}{\kappa_1\Omega_{SD}}\right)\Psi_2 (N_t, \mu, \beta_1 ) \\
				\end{cases}
		\end{align*}

		Therefore, substituting $\ell_A$ into \eqref{eq_ell_2} and combining with \eqref{eq_ell_1}, after some basic mathematical manipulation, we obtain the final result in \eqref{eq_pso}, thus completing the proof.

	\section{Proof of Corollary \ref{corollary:pnzsc}} \label{appendix: B}
		From the definition of the probability of non-zero secrecy capacity given in \eqref{eq_Pnzsc_defi}, by applying the total probability theorem, $P_{nzsc}^{AnJ} $ can be expressed as
		\begin{align*}
			P_{nzsc}^{AnJ} = & \underbrace{ \Pr \{ C_s > 0 | \Phi = \Phi_d \} \Pr \{\Phi = \Phi_d \} }_{\ell_3}  \\
			& + \underbrace{ \Pr \{ C_s >0 | \Phi = \Phi_o \} \Pr \{\Phi = \Phi_o \} }_{\ell_4}  \numberthis
		\end{align*}
		Again, as no secret is transmitted in DEH mode, $\Pr \{ C_s > 0 | \Phi = \Phi_d \} = 0$. Therefore, we have $\ell_3= 0$. And similar to \eqref{eq_ell_2}, we can recast $\ell_4$ as
		\begin{align*} \label{eq_ell_4}
			\ell_4
				& = \Pr \{ (C_s > 0) \cap (C_{SD} \ge R_s) \cap (\varepsilon[k] \ge E_{th}) \} \\
				& = \underbrace{\Pr \{ (C_s > 0) \cap (C_{SD} \ge R_s)\} }_{\ell_{B}} \sum_{i=\tau}^L \xi_{FD,i} \numberthis
		\end{align*}
		Substituting \eqref{eq_channel_condition}, \eqref{eq_Pnzsc_defi} and \eqref{eq_SINR_D} into \eqref{eq_ell_4}, and performing basic mathematical manipulations, we obtain
		\begin{align*} \label{eq_ell_B}
			& \ell_B \\
			& = \Pr \left\{\left( H_{SD} > \frac{ \gamma_E }{\kappa_1} \right) \bigcap \left( H_{SD} \ge \frac{2^{R_s}-1}{\kappa_2} \right) \right\} \\
			& = \Pr \left\{  \left( H_{SD} > \frac{ \gamma_E }{\kappa_1} \right) \bigcap \left( \frac{\gamma_E}{\kappa_1} \ge \frac{2^{R_s}-1}{\kappa_2} \right) \right\} \\
			& \quad + \Pr \left\{  \left( H_{SD} \ge \frac{2^{R_s}-1}{\kappa_2} \right) \bigcap \left( \frac{\gamma_E}{\kappa_1} < \frac{2^{R_s}-1}{\kappa_2} \right) \right\} \\
			& = \displaystyle \int_{ \frac{\kappa_1 \left(2^{R_s}-1\right) }{\kappa_2} }^\infty \int_{\frac{\gamma_1}{\kappa_1}}^{\infty} f_{H_{SD}}(H_{SD}) \! f_{\gamma_E}(\gamma_E) \mathrm{d}H_{SD} \; \mathrm{d} \gamma_E \\
			& \quad + \displaystyle \int_{0}^{\frac{\kappa_1 \left(2^{R_s}-1\right) }{\kappa_2}} \int_{\frac{\left(2^{R_s}-1\right)}{\kappa_2}}^{\infty} f_{H_{SD}}(H_{SD}) \! f_{\gamma_E}(\gamma_E) \mathrm{d}H_{SD} \; \mathrm{d} \gamma_E \numberthis
		\end{align*}
		Substituting \eqref{eq_CDF_HSD} together with \eqref{eq_CDF_rE} and \eqref{eq_PDF_rE} into \eqref{eq_ell_B}, and applying \cite[Eq. (3.352.4) and (3.353.2)]{gradshtein_table_2007} to solve the resultant integrals, we derive $\ell_B$ as
		\begin{align*} \label{eq_ell_B}
				\ell_B
				= \begin{cases}
				\Big( \frac{\sigma_E^2 }{P_S\Omega_{SE}} \Psi_1(1, \mu_2, \beta_1+\beta_2) & \text{if } N_t = 2 \\
					\quad + \Psi_1 (2, \mu_2, \beta_1+\beta_2 ) \Big) \\
					\quad \times \varphi^{-1} e^{- \beta_2\mu_2} \\
					\quad + \exp\left(- \frac{2^{R_s}-1}{\kappa_2 \Omega_{SD}} \right) F_{\gamma_E}(\beta_2) \\
				\\ %
				 \Big( \frac{\sigma_E^2 }{P_S\Omega_{SE}} \Psi_2 (N_t-1, \mu_2, \beta_1+\beta_2 ) & \text{if } N_t \ge 3 \\
					\quad + (N_t-1) \Psi_2 (N_t, \mu_2, \beta_1+\beta_2 ) \Big)    \\
					\quad \times \beta_1^{N_t-1} \, e^{- \beta_2\mu_2 } \\
					\quad + \exp\left(- \frac{2^{R_s}-1}{\kappa_2 \Omega_{SD}} \right) F_{\gamma_E}(\beta_2)   \numberthis
 				\end{cases}
		\end{align*}
		Therefore, substituting $\ell_B$ into \eqref{eq_ell_4}, after some basic mathematical manipulation, we obtain the final result in \eqref{eq_p_nzsc}, thus completing the proof.	

	\section*{Acknowledgment}
	The authors would like to thank the editor 
	and the anonymous reviewers for their insightful comments and suggestions that greatly help improve the quality of this paper.

	\ifCLASSOPTIONcaptionsoff \newpage \fi

	
	
	\bibliographystyle{IEEEtran}
	\bibliography{IEEEabrv,bibli}

\end{document}